\documentclass[manuscript]{emulateapj}

\usepackage{multirow}
\usepackage{graphicx}
\usepackage{amsmath, amsthm, amssymb}
\usepackage{color}

\newcommand{\sech}{{\rm sech}}

\begin{document}

\title{Core and filament formation in magnetized, self-gravitating isothermal layers.}
\author{Sven Van Loo, Eric Keto, Qizhou Zhang}
\affil{Harvard-Smithsonian Center for Astrophysics, 60 Garden Street, Cambridge, MA 02138, USA}
\email{svanloo@cfa.harvard.edu}

\begin{abstract}
We examine the role of the gravitational instability in an isothermal, self-gravitating layer 
threaded by magnetic fields on the formation of filaments and dense cores. Using numerical 
simulation we follow the non-linear evolution of a perturbed equilibrium layer. The linear 
evolution of such a layer is described in the analytic work of \citet[][]{Nagaietal1998}. We 
find that filaments and dense cores form simultaneously. Depending on the initial magnetic 
field, the resulting filaments form either a spiderweb-like network (for weak magnetic fields) 
or a network of parallel filaments aligned perpendicular to the magnetic field lines (for strong 
magnetic fields). Although the filaments are radially collapsing, the density profile of their 
central region (up to the thermal scale height) can be approximated by a hydrodynamical 
equilibrium density structure. Thus, the magnetic field does not play a significant role in 
setting the density distribution of the filaments.  The density distribution outside of the 
central region deviates from the equilibrium. The radial column density distribution is then 
flatter than the expected power law of $r^{-4}$ and similar to filament profiles observed with 
{\it Herschel}. Our results does not explain the near constant filament width of $\sim 0.1$pc. 
However, our model does not include turbulent motions. It is expected that accretion-driven 
amplification of these turbulent motions provides additional support within the filaments against 
gravitational collapse.  Finally, we interpret the filamentary network of the massive star 
forming complex G14.225-0.506 in terms of the gravitational instability model and find that the 
properties of the complex are consistent with being formed out of an unstable layer threaded by 
a strong, parallel magnetic field.
\end{abstract}


\maketitle

\section{Introduction}
Molecular clouds exhibit a hierarchical density structure with stars forming in the densest 
regions. Often, these star-forming complexes have an elongated, filamentary shape 
\citep[e.g.][]{SchneiderElmegreen1979,JohnstoneBally1999, Mizunoetal1995, Goldsmithetal2008}. 
In fact, recent observations by the {\it Herschel} Space Observatory show that parsec-scale 
filaments are ubiquitous in the interstellar medium (ISM) \citep[e.g.][]{Andreetal2010, 
Menshchikovetal2010, Arzoumanianetal2011, Molinarietal2011}. The filaments tend to extend out 
from dense star-forming hubs \citep[][]{Myers2009,Liuetal2012,GalvanMadridetal2013} and, 
sometimes, these hub-filaments structures are part of a larger-scale pattern with parallel 
filaments \citep[][]{Busquetetal2013}.

The filaments exhibit a nearly universal width of $\sim 0.1$ pc (measured by FWHM of the column 
density), independent of the central column density or length of the filament 
\citep[][]{Arzoumanianetal2011, Palmeirimetal2013}. Furthermore, the radial-dependence of the 
column density profile is flatter than expected for an isothermal cylindrical filament in 
hydrostatic equilibrium \citep[][]{Ostriker1964}.  A number of explanations have been suggested 
for the observed density profiles: the filaments correspond to stagnant gas in locally colliding 
flows \citep[][]{Perettoetal2013}, they are isothermal equilibrium cylinders in pressure-balance 
with the external medium \citep[i.e. a radially-truncated cylinder in hydrostatic 
equilibrium;][]{FischeraMartin2012, Heitsch2013}, or are supported by magnetic fields 
\citep[][]{FiegePudritz2000}. 

Thus, the origin of the filamentary cloud structure still remains unclear
and heavily debated. 
The clouds can be 
formed by compressive motions of gravitational and/or turbulent nature 
\citep[][]{Ballesterosetal2007}. Indeed, many numerical simulations produce filamentary 
structures by either isothermal driven turbulence \citep[e.g.][]{PadoanNordlund2002, 
deAvillezBreitschwerdt2005,BallesterosMacLow2002}, thermal instability in large-scale convergent 
flows \citep[e.g.][]{Vazquezetal2003, AuditHennebelle2005, Heitschetal2008} or behind a shock 
front \citep[e.g.][]{KoyamaInutsuka2000,VanLooetal2007,VanLooetal2010}, or by gravitational
instabilities in self-gravitating sheets \citep[e.g.][]{NakajimaHanawa1996, Umekawaetal1999}. 

Some
of the observational evidence points to a turbulent formation process for filaments and 
dense cores in the ISM.  However, observations often show cores within clouds with linewidths 
that require little additional support beyond thermal pressure \citep[e.g.][]{Myers1983, 
Ketoetal2004}. These quiescent cores are adequately modelled by Bonnor-Ebert spheres confining 
the cores by self-gravity and external pressure \citep[][]{Bonnor1956,Alvesetal2001,
Tafallaetal2004}. Such objects show that gravity plays a more prominent role than assumed in 
the turbulent models.

Magnetic fields are also important in the formation of the filaments. Studies comparing 
the orientation of the magnetic field with respect to the diffuse gas provide more insight 
\citep[e.g.][]{Li_etal_2013}. Optical polarization measurements of the Taurus molecular cloud 
show that the magnetic field in the diffuse gas is orientated perpendicular to the axis of 
the B216 and B217 filaments \citep[][]{Goodmanetal1992}. This suggests that the filaments form 
as gas streams along the magnetic field lines \citep[e.g.][]{Ballesterosetal1999}.  However, 
the long axis of the L1506 filament in Taurus and of the $\rho$ Ophiuchus clouds lies along 
the magnetic field \citep[][]{Goldsmithetal2008,Goodmanetal1990}.

In this paper we focus on the role of self-gravity and magnetic fields on the formation 
of filaments and cores within self-gravitating layers. We neglect the effect of turbulent
motions (other than produced by gravitational and magnetic instabilities).  Self-gravitating 
isothermal layers are unstable to perturbations if the exciting modes  are of sufficiently 
long wavelength \citep[][]{Ledoux1951} and they fragment into clumps or thin filaments 
\citep[][]{Miyamaetal1987}. When the layers are threaded by magnetic fields, fragmentation still 
occurs. Using a linear perturbation analysis, \citet[][]{Nagaietal1998} show that, for a layer
with a thickness larger than the pressure scale height, the magnetic field suppresses the 
growth of perturbations perpendicular to the field. Perturbations along the field are unaffected 
so that parallel filaments form within the slab. On the other hand, for a layer with a thickness 
smaller than the scale height, filaments form with their axis orientated along the magnetic 
field.  The separation of the filaments is given by the wavelength of the fastest growing mode.  
The filaments then condense into clumps and dense cores as long as the filaments are in a 
quasi-static equilibrium \citep[][]{InutsukaMiyama1992}. 

We extend the study of \citet[][]{Nagaietal1998} into the non-linear regime and discuss the 
resulting structures. In Sect.~\ref{sect:initial_conditions} we describe the initial 
conditions of the self-gravitating layer and the numerical model used. Then, we discuss 
the fragmentation of the layer in absence (Sect.~\ref{sect:no_magnetic_field}) and presence 
of a parallel or perpendicular magnetic field (Sect.~\ref{sect:magnetic_field} and 
\ref{sect:perp_magnetic_field}). In Sect.~\ref{sect:comparison} we interpret the structure of
the molecular cloud G14.225-0.506 in terms of the discussed gravitational instability model. 
Finally, we conclude and discuss our results in Sect.~\ref{sect:discussion_and_conclusions}.

\section{Model set-up}\label{sect:initial_conditions}
\subsection{Initial conditions}
For our initial conditions we assume a self-gravitating isothermal layer that extends to 
infinity along the $x,y$-plane. The layer is initially in hydrostatic equilibrium, so that 
the density distribution along the $z$-axis is given by \citep[][]{Spitzer1942} 
\begin{equation}\label{eq:Spitzer}
   \rho(z) = \rho_0\ \sech^2(z/H),
\end{equation}
with $H= a/\sqrt{2\pi G\rho_0}$ the pressure scale height, $a$ the isothermal sound speed,
$G$ the gravitational constant and  $\rho_0$ the unperturbed density at the midplane 
(i.e. at $z=0$). The layer is truncated at $\pm z_B$ and confined by a constant external 
thermal gas pressure $p_{\rm g, ext} = a^2 \rho(z_B)$. In this paper we will not study the 
effect of the external pressure on fragmentation. Therefore, we only consider a layer bounded 
by a low external pressure, i.e. $p_{\rm g, ext} \ll p_{\rm g,0}$. We adopt $z_B = 2H$, so 
that $p_{\rm ext} = 0.071 p_{\rm g,0}$.

Outside the layer, not only the pressure, but also the density, $\rho_{\rm ext}$, is kept 
constant at a 
small 
fraction of the density at the boundary (see later for exact value). 
This means that the external temperature differs from the internal temperature. We, therefore, modify 
the isothermal equation of state to be 
\begin{equation}
   p_{\rm g} = \rho \left((1 - \alpha) \frac{p_{\rm ext}}{\rho_{\rm ext}} + \alpha a^2\right).
\end{equation}
This reflects the difference in external and internal temperature with $\alpha$ a scalar 
that is 1 inside the layer and 0 outside. 

The layer 
and the external medium are
threaded by a uniform magnetic field. For parallel models the magnetic field is 
along the $x$-axis, while, for perpendicular models, it is along the $z$-axis.  The field 
strength is given by $\beta = p_m/p_g$ at the midplane with $p_m = B^2/2$ the magnetic pressure. 
For our models we use 
$\beta = \infty {\rm \ (i.e.\ no\ magnetic\ field)}, 10, 1 {\rm \ and}\ 0.1$. For simplicity, 
our simulations are performed dimensionless. We adopt $\rho_0 = 2\pi$, $G = 1$ and $a = 1$ so 
that $H = 1/2\pi$ and the dynamical time  $t_{\rm dyn} = 1/\pi$. The free-fall time associated 
with gas at the midplane is $\sim 1/5$. For a layer with a central density of $1000~{\rm cm^{-3}}$ 
(and thus a surface density of $\approx 10^{21}~{\rm cm^{-2}}$) and a temperature of 10~K, these 
values correspond to $H = 0.159$~pc, $B = 27~{\rm \mu G}$ (for $\beta = 1.0$) and 
$t_{\rm dyn} = 1.57$~Myr. In the rest of the paper we will evaluate the dimensionless results in 
terms of this layer. 

This set-up is similar to the one used in the linear analysis of \citet[][]{Nagaietal1998} 
allowing a comparison between analytically and numerically derived growth rates of the 
gravitational instability. To study the gravitational instability of a pressure-confined 
self-gravitating layer, we perturb the equilibrium state. We do this in two different ways, 
i.e. we superimpose the density either with a sinusoidal density perturbation of a given 
wavelength or with randomly generated noise. The former is used to determine the growth 
rates to compare with analytic values, while the latter shows how different wave modes 
interact with each other. From the linear analysis \citet[][]{Nagaietal1998} suggest 
that first the filaments form and only then the dense cores.

\subsection{Numerical code and domain}
To solve the ideal magnetohydrodynamics (MHD) equations we use the Adaptive Mesh Refinement 
(AMR) MHD code MG \citep[][]{VanLooetal2006,Falleetal2012}. The basic algorithm is a 
second-order Godunov scheme with a linear Riemann solver \citep[][]{Falle1991}. To ensure that 
the solenoidal constraint is met, a divergence cleaning algorithm is implemented in the 
numerical scheme \citep[][]{Dedneretal2002}.  A hierarchy of grids levels is used with a mesh 
spacing on grid level $n$ of $\Delta x/2^n$ with $\Delta x$ the cell size at the coarsest level.  
While the two coarsest levels cover the entire domain, finer levels generally do not.  
Refinement is on a cell-by-cell basis and is controlled by error estimates based on the 
difference between solutions on different grid levels. Self-gravity is computed using a full 
approximation multigrid to solve the Poisson equation.

The numerical domain is given by $-2~<~z~<~2$ and $-L/2~<~x,y~< L/2$ where $L$ is either  
$\lambda_{crit}/2, \lambda_{\rm crit}, \lambda_{\rm max}\ {\rm or}\ 4\lambda_{\rm max}$. 
Here, $\lambda_{\rm crit} = 2\pi H$ is the critical wavelength below which the gravitational 
instability is suppressed and $\lambda_{\rm max} = 4\pi H$ the wavelength for which the growth 
rate is maximal \citep[see][]{Nagaietal1998}. For our adopted values, 
$\lambda_{\rm crit} \approx  1$ and $\lambda_{\rm max} \approx 2$. As the layer is assumed to 
be infinite, we set the boundary condition on both the $x$ and $y$-axis to periodic. For the 
$z$-axis we use free-flow boundary conditions.  

The resolution of the simulation is set by the vertical extent of the slab. It is necessary
to properly resolve the pressure scale height $H$ in order to find a proper balance between 
pressure gradients and self-gravity. We resolve the scale height by at least 5 cells.  Therefore, 
we use an AMR mesh with the coarse level having 4 grid cells along the axis with the shortest 
length (along the $x,y$-axis for the runs up to $L = \lambda_{\rm max}$ and along the $z$-axis 
for $L = 4\lambda_{\rm max}$) and 128 cells along the $z$-axis for the finest grid level. This 
means that the number of additional grid levels varies with different models, i.e. from 2 extra 
grid levels for $L = \lambda_{\rm crit}/2$ up to 5 grid levels for $L= 4\lambda_{\rm max}$. This 
resolution is more than sufficient to resolve the Jeans' length as suggested by 
\citet[][]{Trueloveetal1997} to avoid artificial fragmentation.  In fact, artificial fragmentation 
is only expected above $\rho > \pi a^2/16 G \Delta x^2 \approx 200$.

As we are interested in the gravitational instability in the equilibrium layer, we need to 
ensure that the external medium does not affect the gravitational instability in the layer 
in any way. By assuming a constant pressure in the external medium, the external medium 
needs to be a vacuum. (The gravitational force is zero if there is no pressure force.)
However, in a grid code, we cannot set the density to zero. For a non-zero external density,
the self-gravity of the layer drags the external medium towards it. Then the external medium 
affects the hydrostatic layer because gas is accreted onto the layer and because the ram pressure 
contributes to the total external pressure. We find that the external 
density needs to be at most $10^{-4}\ \rho(z_{\rm B})$ not to affect the equilibrium layer.
As the external pressure is $\rho(z_{\rm B}) a^2$, this means that the external sound speed 
is $10^2$ times larger than in the layer. This significantly limits the stable numerical 
time step derived from the Courant condition $\approx \Delta x/a$. Because of this constraint, 
we use a different approach, i.e. we artificially switch off 
the self-gravity in the external medium. This is easily done as the different media have a 
different value of the scalar $\alpha$. This intervention allows us to set the external medium 
density to $0.1\ \rho(z_{\rm B})$ without affecting the equilibrium layer and gives us a speed up
of more than a factor of ten in computational time.


\section{No magnetic field}\label{sect:no_magnetic_field}
In order to determine the effect of the magnetic field on the density distribution and 
geometry of the filaments, we first study the onset and evolution of the gravitational 
instability in the absence of a magnetic field. We start by confirming the 
theoretically derived growth rates and continue by examining the interaction of 
different wave modes.

\begin{figure}
\begin{center}
\includegraphics[width=8cm]{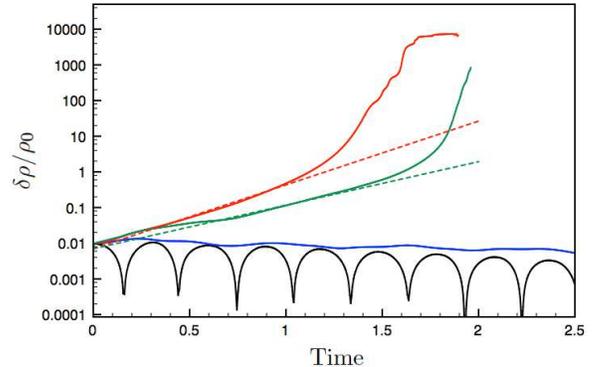}
\caption{The temporal evolution of the normalized density perturbation amplitude
for a wave mode with a wavelength $\lambda_{\rm crit}/2$ (black), $\lambda_{\rm crit}$ (blue), 
$\lambda_{\rm max}$ (red) and $4 \lambda_{\rm max}$ (green). The dashed lines show the linear 
evolution using the theoretically derived growth rates. For a layer with a central density 
of $1000~{\rm cm^{-3}}$, i.e. our default layer, a unit of time is $\sim 4.9$Myr.}
\label{fig:growth_rates_hydro}
\end{center}
\end{figure}

\paragraph{Growth rates} Figure~\ref{fig:growth_rates_hydro} shows the evolution of the 
maximum density perturbation amplitude for modes with a wavelength of $\lambda_{\rm crit}/2$, 
$\lambda_{\rm crit}$, $\lambda_{\rm max}$ and $4 \lambda_{\rm max}$. The initial amplitude 
of the perturbation is 1\% of the local density. For a perturbation with a wavelength below 
$\lambda_{\rm crit}$, the amplitude of the perturbation gradually decreases. The wave mode 
is damped and the self-gravitating layer is stable under such a perturbation. This remains 
so until the wavelength of the perturbing mode reaches $\lambda_{\rm crit}$. This critical 
wavelength (hence $\lambda_{\rm crit}$) has a zero growth rate and the amplitude of the 
critical wave mode remains roughly constant for multiple dynamical time scales. It also 
marks the transition to a layer that is unstable to perturbations with 
$\lambda > \lambda_{\rm crit}$.  Our simulations reproduce the theoretically derived linear 
growth rates of \citet[][]{Nagaietal1998} for these longer wavelength perturbation very 
well. During the linear growth phase, the amplitude of the perturbation grows as 
$\sim \exp(\omega t)$. For the fastest growing wave mode with wavelength $\lambda_{\rm max}$, 
we find $\omega \approx 4.12$, which is close to the analytically derived value of 
$\omega \approx 4.49$ \citep[][]{ElmegreenElmegreen1978}. For $4\lambda_{\rm max}$, 
$\omega \approx 2.81$ and is similar to the value inferred from Fig.~1 of 
\citet[][]{Nagaietal1998}.  Note that the linear growth phase ends when 
$\delta\rho \approx \rho_0$ and the growth becomes non-linear. This non-linear growth phase 
lasts until nearly all the gas of the layer is concentrated in a single dense filament which 
is starting to fragment into cores. We do not have sufficient resolution to follow this 
fragmentation and, at this time, we end the simulation.

\begin{figure}
\begin{center}
\includegraphics[width=8cm]{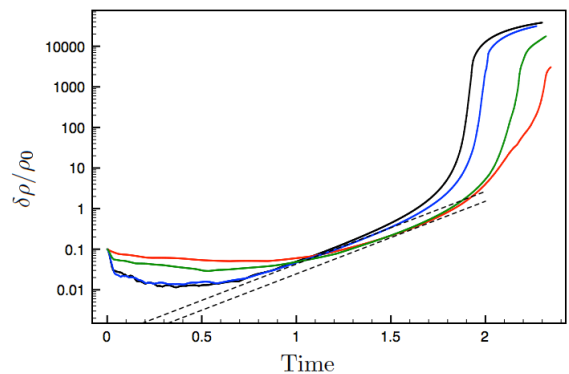}
\caption{The temporal evolution of the normalized density perturbation (solid line) 
when random density perturbations are imposed on the equilibrium slab  for a hydrodynamical 
(black), $\beta = 10$ (blue), 1 (green) and 0.1 (red) model. The dashed lines shows the 
linear growth of a wave mode with a wavelength $\lambda_{\rm max}$. For our default layer, 
unit of time is $\sim 4.9$Myr.}
\label{fig:growth_rate_noise}
\end{center}
\end{figure}

\paragraph{Filament and core formation} We have confirmed the growth rates for different 
wave modes, but we need to understand how the different wave modes interact with each other. 
Therefore, we superimpose the equilibrium density with random noise. The maximal amplitude of 
the density noise is 10\% of the local density. We allow wave modes with a wavelength up to 
$4\lambda_{\rm max}$ to develop within the layer, i.e. the numerical domain is 
$-2\lambda_{\rm max} < x,y < 2\lambda_{\rm max}$. Figure~\ref{fig:growth_rate_noise} shows 
the maximum amplitude of the density perturbations. Initially, as we impose random noise
at the grid level, the wave modes have a wavelength below $\lambda_{\rm crit}$. These modes 
are stable and the amplitude thus decreases. However, wave modes with longer wavelengths 
for which the layer is unstable are excited and gradually grow. The fastest growing unstable
wave mode, i.e. the mode with a wavelength $\lambda_{\rm max}$, becomes dominant. This can be 
clearly seen in Fig.~\ref{fig:growth_rate_noise} as the evolution of the maximum perturbation
amplitude can be described by the linear growth phase of that wave mode.

\begin{figure}
\begin{center}
\includegraphics[width=8cm]{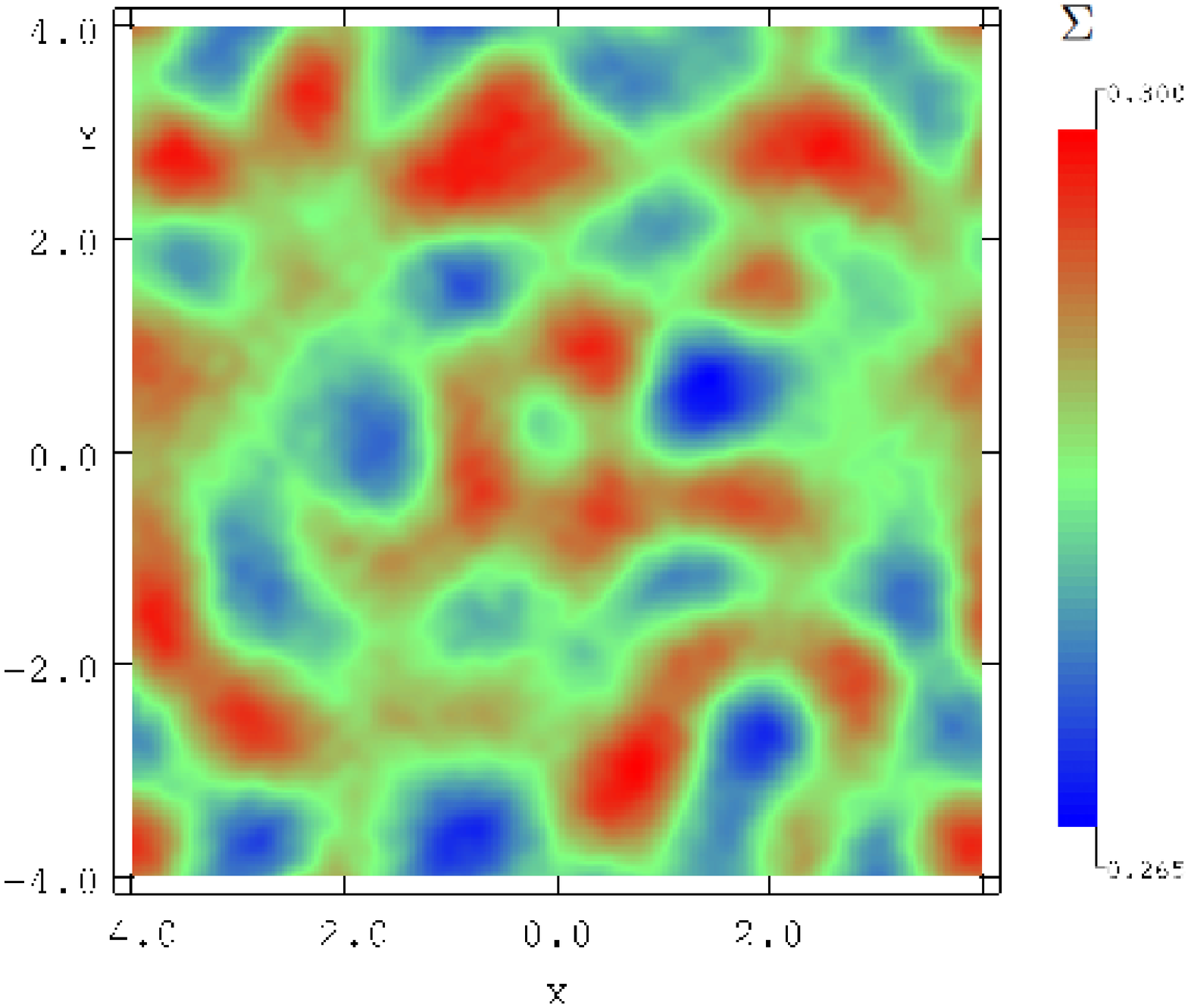}
\includegraphics[width=8cm]{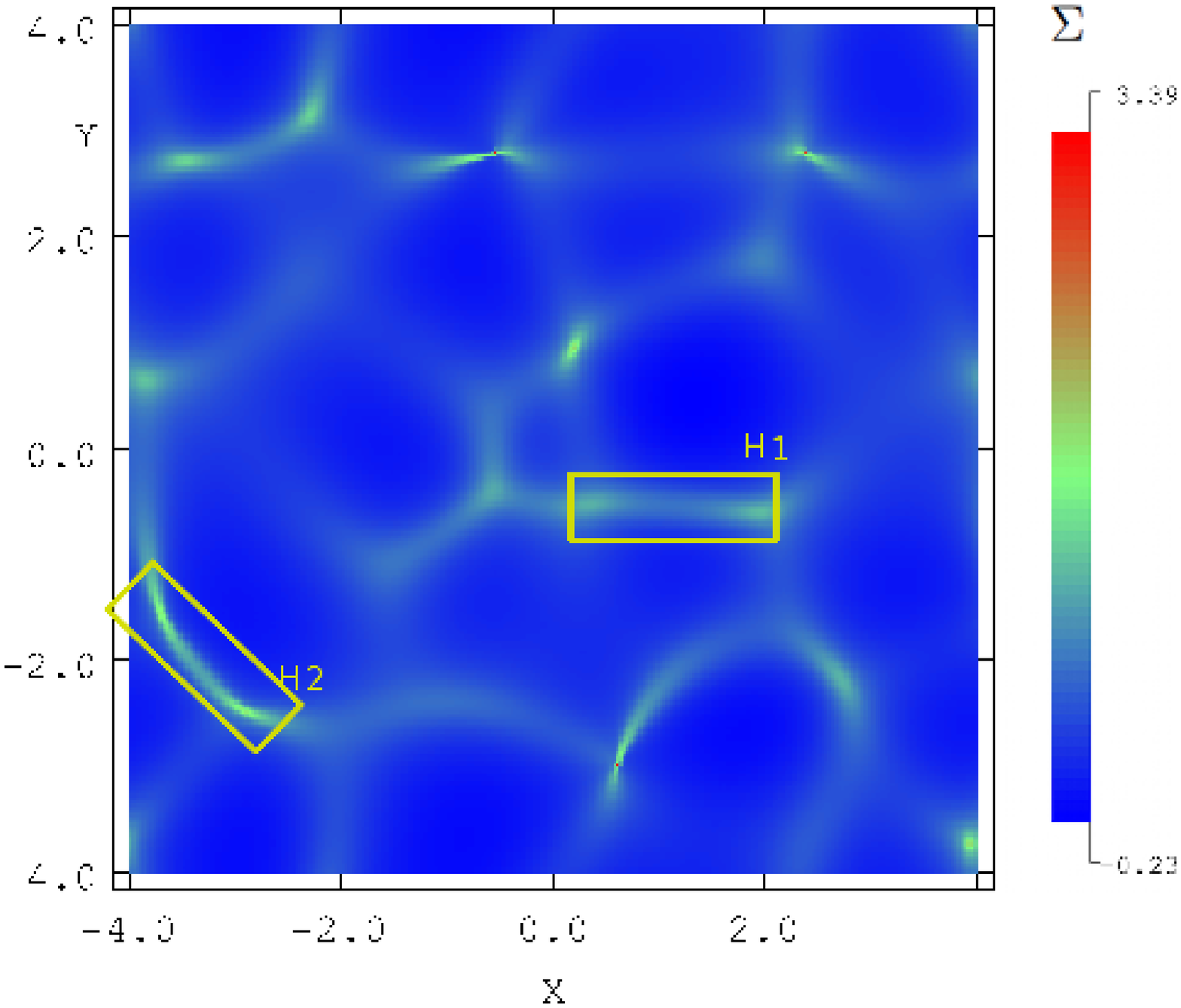}
\caption{The logarithmic surface density along the $z$-axis for a layer without magnetic fields
at $t = 1$ (top) and at $t = 2$ (bottom). The boxes show filaments selected for further analysis.
In terms of our default layer, the area shown is 8$\times$8~pc$^2$ and the unit of surface 
density is 5$\times$10$^{20}~{\rm cm^{-2}}$. 
}
\label{fig:column_pl_hydro}
\end{center}
\end{figure}

\begin{figure}
\begin{center}
\includegraphics[width=8cm]{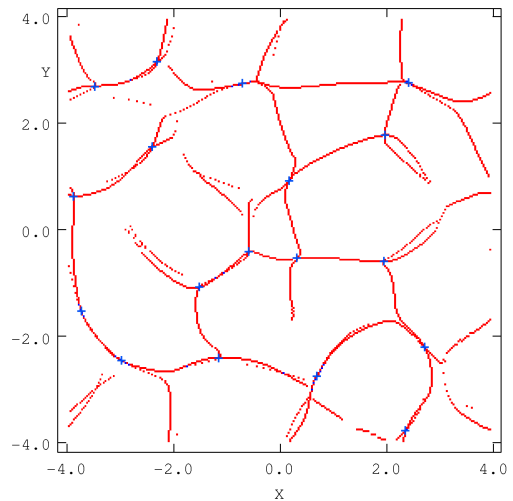}
\caption{Ridge identification for the column density of the bottom panel in 
Fig.~\ref{fig:column_pl_hydro}.  The local maxima are indicated by blue crosses.}
\label{fig:ridge_pl_hydro}
\end{center}
\end{figure}

Figure~\ref{fig:column_pl_hydro} shows the resulting surface density integrated along the 
$z$-axis at two different times, i.e. at the beginning of the linear growth phase at 
$t = 1$ and near the end of the non-linear growth phase at $t = 2$. A network of 
interconnecting filaments with embedded cores has formed. Note that regions of high column 
density are already present early on in the linear phase. In fact, there is a one-to-one 
correlation.  To better understand the interplay between the different filaments and the 
positions of the cores, we use a ridge finding algorithm to expose the skeleton of the 
network and to locate column density maxima \citep[e.g.][]{Lindeberg1998}.  The ridges and 
local maxima for the bottom panel of Fig.~\ref{fig:column_pl_hydro} are shown in 
Fig.~\ref{fig:ridge_pl_hydro}. Note that filaments have a separation of about 
$\lambda_{\rm max}$ indicating that the filaments are indeed formed by the most unstable 
wavelength (as already inferred from the growth rates). Although, in the absence of magnetic 
fields, the most unstable wave mode has the same growth rate in all directions, filaments 
predominantly form along the $x$ and $y$ axes. This is most likely because the flow equations 
are evolved along the Cartesian coordinate axes in the numerical code and thus excite the 
wave modes along these directions. Another interesting feature is that the local maxima, 
i.e. dense cores, lie at the intersections of the filaments. This is reminiscent of the 
hub-filament structures discussed by \citet[][]{Myers2009} and the network of filaments seen 
in the {\it Herschel} observations of the Rosette molecular cloud and Pipe nebula 
\citep[][]{Schneideretal2012,Perettoetal2013}. Such a network is also present in cloud 
formation simulations which include turbulence \citep[e.g.][]{Vazquezetal2003}, but is 
generated due to dynamical (other than gravity) and thermal instabilities.  Other random 
noise initializations produce a similar network of filaments with a separation of about 
$\lambda_{\rm max}$ and dense cores at the intersections of filaments.

\paragraph{Radial profile of filaments} Several authors \citep[e.g.][]{Arzoumanianetal2011, 
Palmeirimetal2013, Kirketal2013} characterise the filaments by mapping the observed 
column densities with Plummer-like profiles. They assume that the underlying density 
profile of the cylindrical filament is given by
\begin{equation}\label{eq:plummer}
    \rho(r) = \frac{\rho_0}{\left[ 1 + \left(r/R_0\right)^2\right]^\frac{p}{2}},
\end{equation}
where $r$ is the cylindrical radius, $R_0$ the characteristic radius of the flat inner region
and $p$ the power-law index of the profile at large radii. For $p = 4$, this profile describes 
an isothermal cylinder in hydrostatic equilibrium with $R_0 = \sqrt{2a^2/\pi G \rho_0}$ the 
scale height \citep[][]{Ostriker1964}.  The resulting column density profile (assuming that the 
filament lies within the plane perpendicular to the integration) is also a Plummer-like profile 
given by
\begin{equation}\label{eq:column_density} 
    \Sigma(l) = \frac{B_p \rho_0 R_0}{\left[ 1 + \left(l/R_0\right)^2\right]^\frac{p-1}{2}},
\end{equation}
where $B_p$ is the Euler beta function with input values $1/2$ and $(p-1)/2$ and $l$ the 
projected radius. 

\begin{figure}
\begin{center}
\includegraphics[width=8cm]{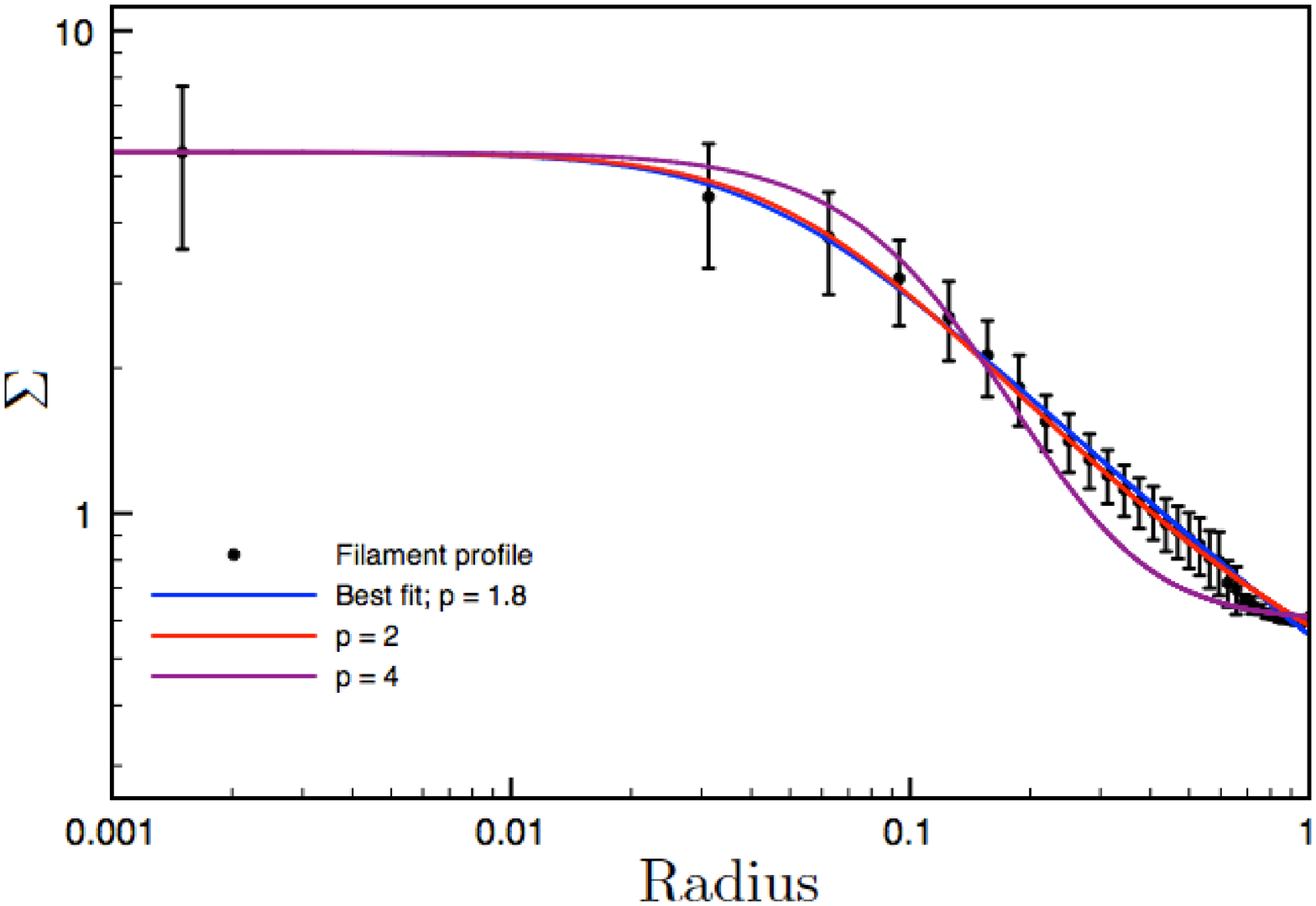}
\includegraphics[width=8cm]{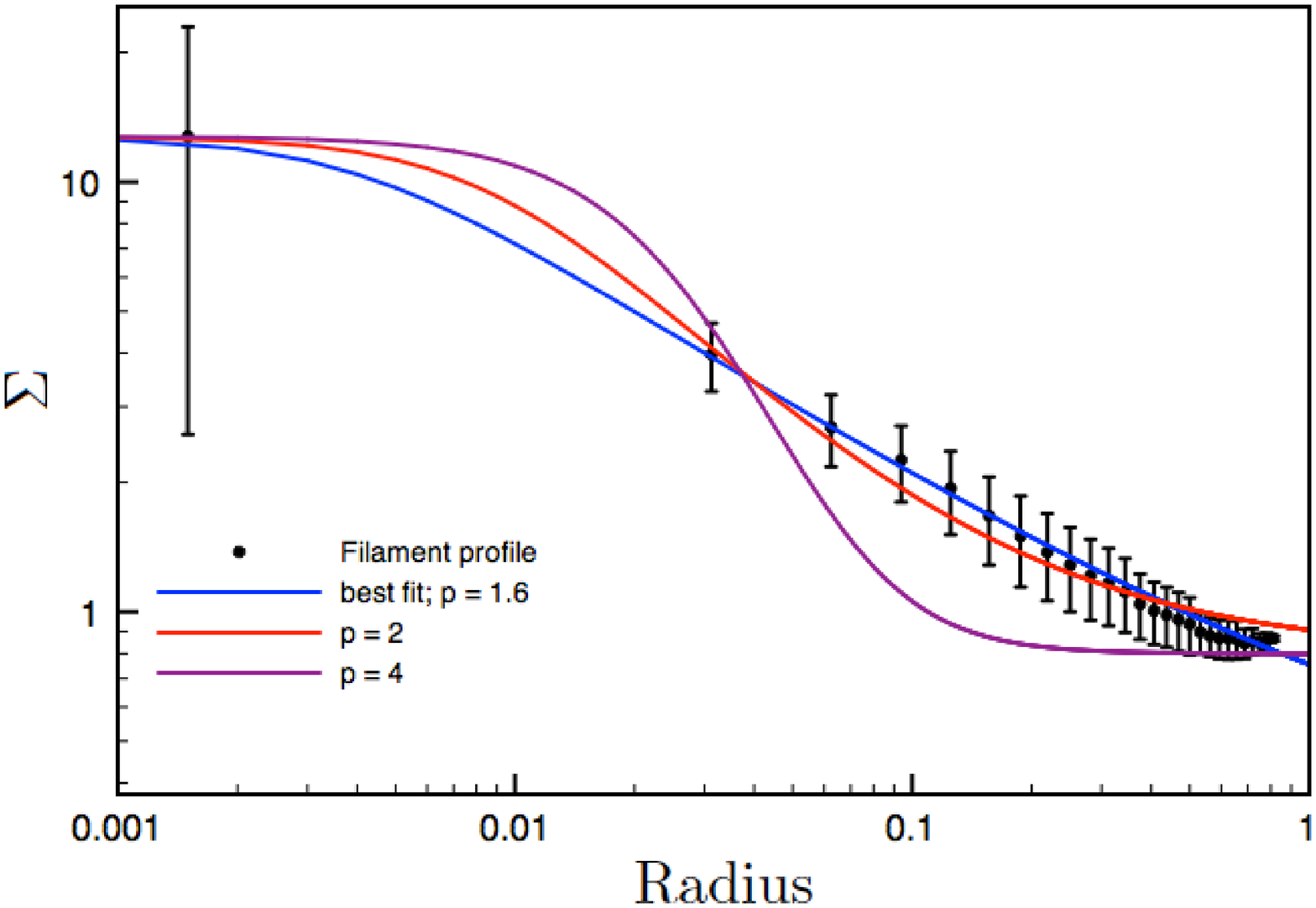}
\caption{Radial column density profile for two selected regions in Fig.~\ref{fig:column_pl_hydro}: 
Region~H1 is on the top panel, while Region~H2 is on the bottom one.  We display 
the central column density at a radius of 0.0015 and not at a radius of 0.0. 
The solid circles show the column densities at a given radius from 
the filament axis averaged over the entire filament while the error bars are the dispersion on the 
mean.  The blue line is the best fit Plummer-profile with a background column density, while the 
red and purple lines show the Plummer-like profile for $p = 2$ and $p=4$, respectively. For our 
default layer, the unit of length is 1~pc and the unit of surface density is 
5$\times 10^{20}~{\rm cm^{-2}}$.}
\label{fig:pl_hydro_profiles}
\end{center}
\end{figure}

We extract several filaments from the column density map  at $t = 2$.
For each filament, we derive the radial distributions perpendicular to the axis at 
each cell along the filament axis.  By combining all the distribution along a filament axis 
we obtain the average radial distribution of the filament. Note that, for curved filaments, 
radial profiles will cross each other, but this will not strongly influence the average radial 
distribution as long as the curvature is small. We then fit the average distribution with the 
theoretical profiles of Eq.~\ref{eq:column_density}. We add a constant 
background column density because the filaments are embedded in a layer.

Figure~\ref{fig:pl_hydro_profiles} shows the profiles and fits for two filaments. 
The best fit for Region~H1 is given by $p = 1.8$, $R_0 = 0.045$ and $\rho_0 = 26.9$, while 
$p = 2$ still produces a good fit. However, an equilibrium model with $p = 4$ cannot reproduce 
the filament profile at all. A similar result is found for Region~H2 where the best fit is 
$p = 1.6$, $R_0 = 0.004$ and $\rho_0 = 683$. These results agree well with observationally 
derived values of $1.5 < p < 2.5$ by \citet[][]{Arzoumanianetal2011} and others. 
The value of $R_0$, however, varies significantly between the different filaments. 
For Region~H1 the inner flat region is about an order of magnitude larger than in Region~H2.

\begin{figure}
\begin{center}
\includegraphics[width=8cm]{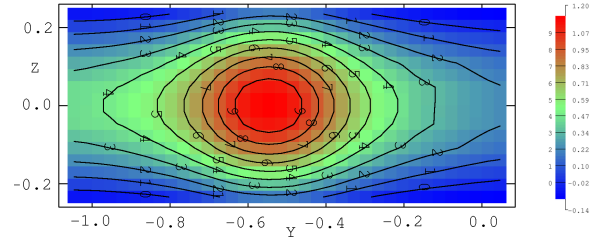}
\caption{
Density cross-section of the filament in Region~H1 of Fig.~\ref{fig:column_pl_hydro} 
at $x = 1.25$ with contour lines. For the default layer, the units of density and length
are 159.2~${\rm cm^{-3}}$ and 1~pc, respectively.}
\label{fig:pl_hydro_crosssection}
\end{center}
\end{figure}

\begin{figure}
\begin{center}
\includegraphics[width=8cm]{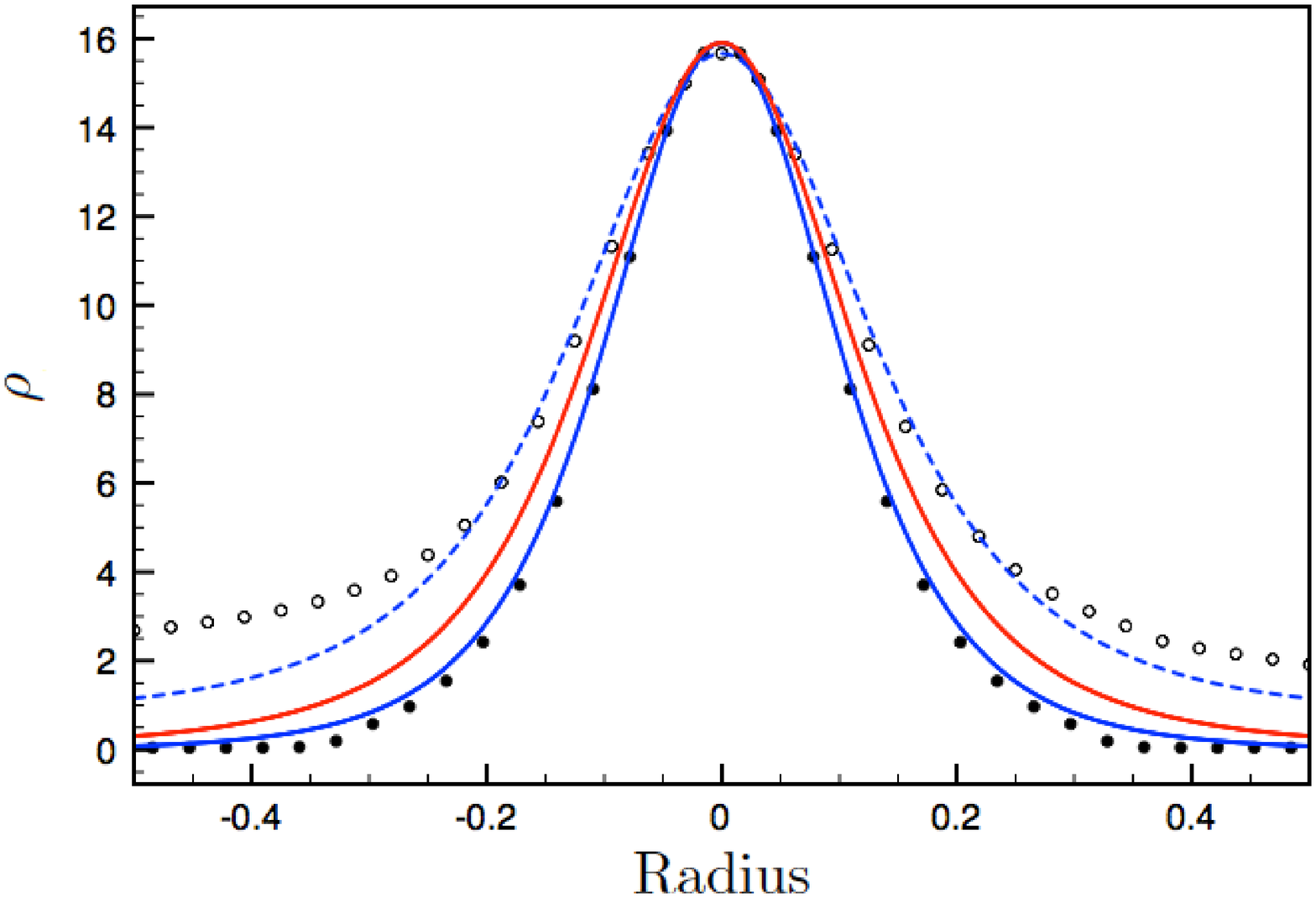}
\includegraphics[width=8cm]{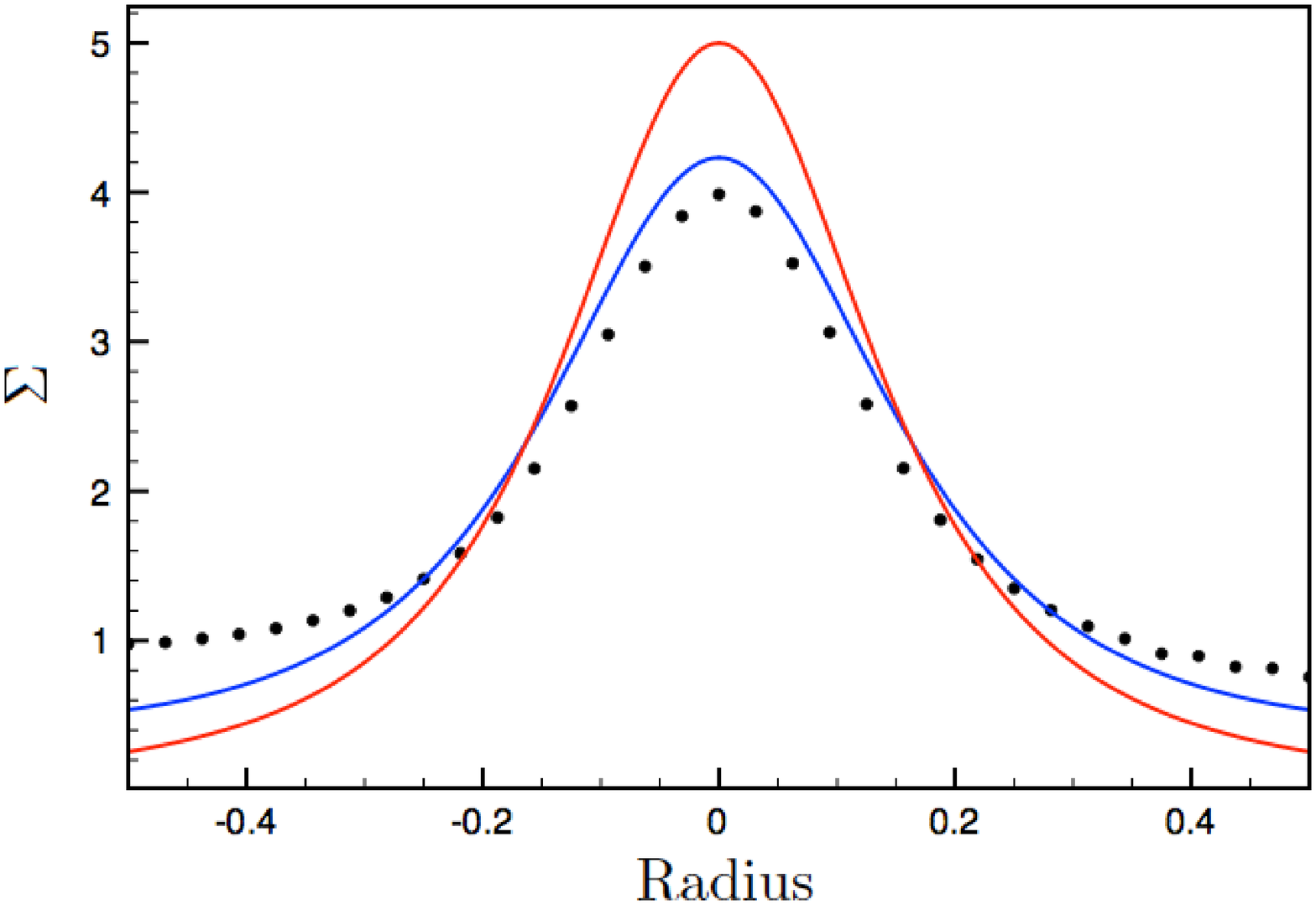}
\caption{
{\it Top:} Density slices along the coordinate axes for Fig.~\ref{fig:pl_hydro_crosssection}. 
The open circles are along the $y$-axis and the solid ones along the $z$-axis. The blue lines 
(solid along $z$-axis and dashed along $y$-axis) are the best fit model for a Schmidt-Burgk profile, 
i.e. $A = 0.6$ and $\rho_c = 3.976$, while the red line is the expected profile for an equilibrium 
cylinder with $\rho_0 = 15.9$. For our default layer, the unit of density is 159.2 ${\rm cm^{-3}}$.
{\it Bottom:} Column density profile of the same figure. The dots are the simulation values,
while the blue, resp. red, solid lines are the expected column density for the Schmidt-Burgk profile,
resp. the Ostriker profile. The unit of surface density corresponds to 5$\times 10^{20}~{\rm cm^{-2}}$,
while the unit of length is 1~pc.}
\label{fig:pl_hydro_crosssection_slices}
\end{center}
\end{figure}

The average radial column density profiles suggest that the filaments are far from
equilibrium. Numerical simulations provide the opportunity to confirm this claim 
directly from the density distribution. We examine different slices across the filament 
in Region~H1. Figure~\ref{fig:pl_hydro_crosssection} shows one of these cross sections. 
Although the center of the filament is nearly axisymmetric, the outer layers 
are flattened along the slab.  An equilibrium profile \citep[][]{Ostriker1964} 
provides a good fit to the central region of the filament, but cannot explain the 
asymmetry along the different axes (see Fig.~\ref{fig:pl_hydro_crosssection_slices}).
This, however, does not imply that the structure is far from equilibrium.  Actually, 
the cross section is reminiscent of a modulated self-gravitating layer in equilibrium 
\citep[][]{Curry2000, Myers2009}. The density distribution of such a modulated layer 
is given by \citep[][]{Schmid-Burgk1967}
\begin{equation}\label{eq:Schmid}
   \rho = \rho_c \frac{1 - A^2}{\left(cosh(z/L) - A\ cos(x/L)\right)^2},
\end{equation}
where $\rho_c$ is the midplane density of the unmodulated layer, $A$ the modulation factor
between 0 and 1 and $L = a/\sqrt{2\pi G\rho_c}$ the scale height. For $A = 0$ we find the 
Spitzer solution for a infinite layer, i.e. Eq.~\ref{eq:Spitzer}, while, for $A \rightarrow 1$,
it converges to the \citet[][]{Ostriker1964} solution for an infinite cylinder. By applying a 
Schmid-Burgk profile to density slices through the center of the cross section shown 
in Fig.~\ref{fig:pl_hydro_crosssection} and along the $y$ and $z$ coordinate 
axes\footnote{The slice along the $z$-axis is not at $z = 0$, but at $z = \Delta x/2$ due to 
the gridding of the numerical domain.}, we find that the density profile of the filament is 
best reproduced with $A = 0.6$ and $\rho_c = 3.976$ (see 
Fig.~\ref{fig:pl_hydro_crosssection_slices}). The fit is good for radii up to the scale 
height $L = 0.2$ above which the density along the midplane is higher.  Other slices of the 
filament can also be described by a Schmid-Burgk profile albeit with somewhat different values 
of $A$ and $\rho_c$ (i.e. $0.55 < A < 0.65$ and $4 < \rho_c < 7.5$).  This variation implies 
a density variation along the filament axis and reflects that cores are embedded within a filament.

By integrating Eq.~\ref{eq:Schmid} along the $z$-axis and using the derived parameters
$A$ and $\rho_c$, we find that the Schmid-Burgk profile reproduces not only the 
density profile, but also the column density profile (see 
Fig.~\ref{fig:pl_hydro_crosssection_slices}). Again the fit is best for radii below the 
scale height, but it exhibits a flatter radial distribution than the 
\citep[][]{Ostriker1964} profile. This partly explains the flatter than the Ostriker 
equilibrium radial profile observed in the mean column density distribution 
(Fig.~\ref{fig:pl_hydro_profiles}). Some of the flattening is due to averaging profiles with 
marginally different parameters.  

\begin{figure}
\begin{center}
\includegraphics[width=8cm]{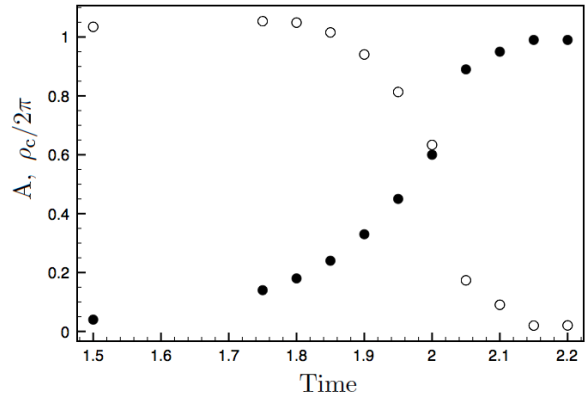}
\caption{Evolution of the filament profile as function of time. The solid circles are the values 
for $A$, while the open ones are the values of $\rho_c$ normalized to the initial value. In terms
of our default layer, a unit of time is 4.9Myr. } 
\label{fig:SBevolution}
\end{center}
\end{figure}

\begin{figure}
\begin{center}
\includegraphics[width=8cm]{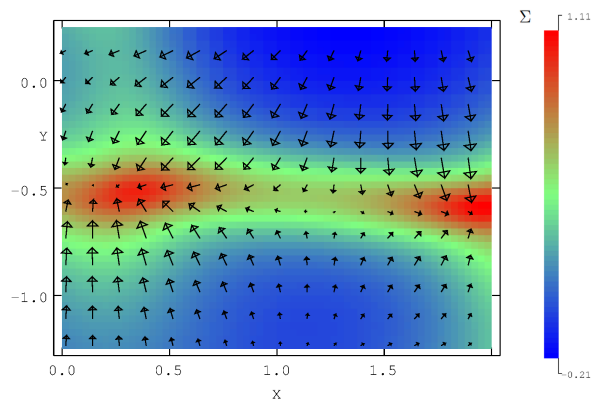}
\caption{Logarithmic column density for Region H1 of Fig.~\ref{fig:column_pl_hydro}. 
The vectors represent the mass-weighted velocities in the plane. Their size is an 
indication of their magnitude with the maximal velocity shown $1.6$ times the sound speed.
Remember that the unit of column density is 5$\times 10^{20}~{\rm cm^{-2}}$ and that 
the sound speed is $\approx 0.2~{\rm km\ s^{-1}}$ in our default layer.} 
\label{fig:region1}
\end{center}
\end{figure}

\paragraph{Evolution of the filaments} Although the density profile shows 
a filament close to equilibrium, it is in fact dynamically evolving. The filament profile 
changes with time, but can, at all instances, be described by a Schmid-Burgk profile 
(see Fig.~\ref{fig:SBevolution}). Initially, the filament is only a small perturbation on 
top of the equilibrium layer (i.e. $A \approx 0$), while it evolves towards an equilibrium 
cylinder at the end of the non-linear phase (i.e. $A \approx 1$).  The evolution of 
a filament thus occurs through a sequence of quasi-equilibrium distributions. Note that 
$\rho_c$ does not remain constant during the evolution.  This suggests that gas does not 
accumulate in this filament segment.  Figure~\ref{fig:region1} indeed shows that, while gas 
flows towards the filaments, it is then diverted towards denser regions (cores) along the 
filament axis.

The evolution of the filaments is set by the mass that the gravitational instability sweeps up.
The expected line mass is essentially determined by the wavelength of the fastest growing 
wavelength and given by $m_{\rm line} \approx \Sigma_0 \lambda_{\rm max}$ with 
$\Sigma_0 \approx 2 \rho_0 H$ the column density of the equilibrium layer. This value 
needs to be compared with the critical value for the line-mass of a stable cylinder, 
i.e. $m_{\rm crit} = 2a^2/G$ \citep[][]{Ostriker1964}. For our model parameters, 
$m_{\rm line} \approx 1.9 m_{\rm crit}$. We can also derive line-masses for different
cross-sections in filaments. For example, we find  $m_{\rm line} \approx 2.6$ for 
Fig.~\ref{fig:pl_hydro_crosssection_slices}). While this is less than the predicted value, 
it still exceed the critical value of $m_{\rm crit} = 2$. It is interesting to note 
that half of the line-mass lies within the central region bound by the scale height. 
Thus, the filament is collapsing radially while maintaining a quasi-equilibrium 
for radii up to the scale height.

\begin{figure}
\begin{center}
\includegraphics[width=8cm]{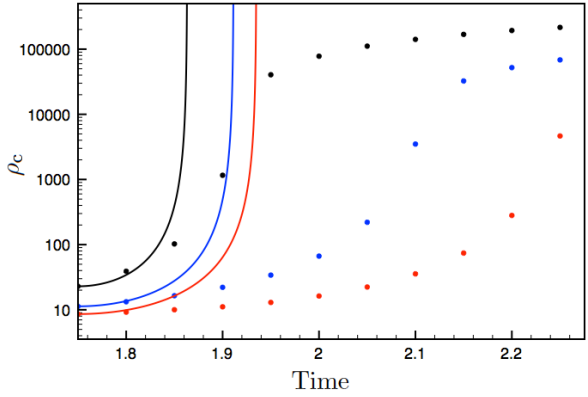}
\caption{The evolution of the central density of three selected cores in the hydrodynamical simulation
(full circles). Each core is represented by a different color (black, blue or red).
The solid lines show the evolution of the central density for homologous collapse 
of a uniform sphere for each core (in the same color). In terms of the default layer, a unit
of time equals 4.9Myr and a unit of density 159.2~cm$^{-3}$. } 
\label{fig:core_evol}
\end{center}
\end{figure}

\paragraph{Evolution of the dense cores} As mentioned earlier, the dense cores arise at the 
junctions of the filaments. The cores are elongated along the axis of the most dense filament. 
As they evolve, the cores become more centrally condensed (see Fig.~\ref{fig:core_evol}). 

The cores collapse at a rate that is slower than expected for pressureless gravitational, 
i.e.  homologous, collapse of a uniform sphere (the free-fall time is between 0.1 and 0.2 for 
all the cores). There are two main reasons. Firstly, the cores are embedded in and are 
formed out of cylindrical filaments. Mass accretion towards the cores is thus 
not isotropic but highly directional. \citet[][]{Toalaetal2012} and \citet[][]{Ponetal2012} 
show that the collapse time scales of non-spherical structures are longer than the corresponding 
spherical free-fall time scale. In the case of a cylindrical cloud, the time-scale depends 
strongly on the aspect ratio of the cloud. 
Note that the above authors only consider the collapse of constant mass structures.
The mass of the filaments does change in our simulation. However, as the line mass of the filament 
is set by the fastest-growing wavelength, the mass of the filaments does not change 
significantly once the evolution is in the non-linear regime.

\begin{figure}
\begin{center}
\includegraphics[width=8cm]{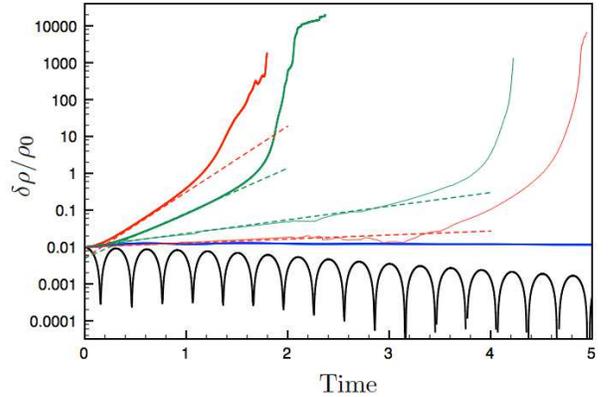}
\caption{The temporal evolution of the normalized density perturbation 
for perturbations with a wavelength $\lambda_{\rm crit}/2$ (black), $\lambda_{\rm crit}$ (blue), 
$\lambda_{\rm max}$ (red) and $4 \lambda_{\rm max}$ (green). The thin solid lines are for 
perturbations along the $y$-axis. The dashed lines show the linear evolution using the 
theoretically derived growth rates. As before, the unit of time corresponds to 4.9Myr for our
default layer.}
\label{fig:growth_rates_blz}
\end{center}
\end{figure}

Secondly, the assumption of pressureless collapse is not valid. The results for the 
filaments show that the inner region can be adequately described by an equilibrium profile in which
self-gravity is nearly balanced by thermal pressure. \citet[][]{InutsukaMiyama1992} show that, 
in the case of equilibrium cylinders,  this significantly reduces the collapse rate. 
As cores form out of filaments, it is likely that the thermal pressure plays a similar role 
in cores. However, we cannot easily test this hypothesis as there is no analytic solution 
for an equilibrium core within a cylindrical filament (and certainly not for a core in a 
Schmid-Burgk like filament). Furthermore, the accretion of gas is not along one filament, 
but multiple ones complicating the expected structure even more.

\begin{figure}
\begin{center}
\includegraphics[width=8cm]{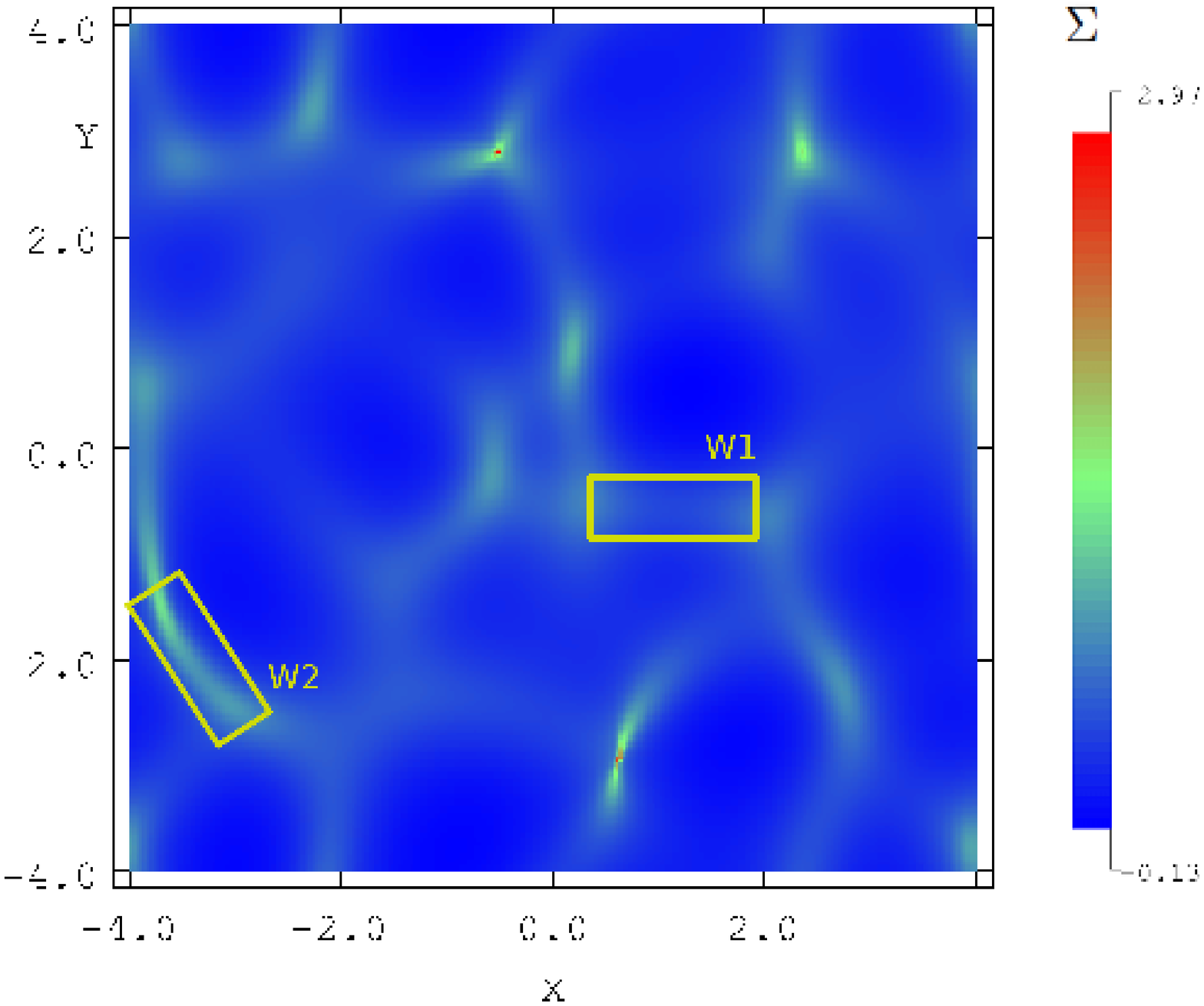}
\includegraphics[width=8cm]{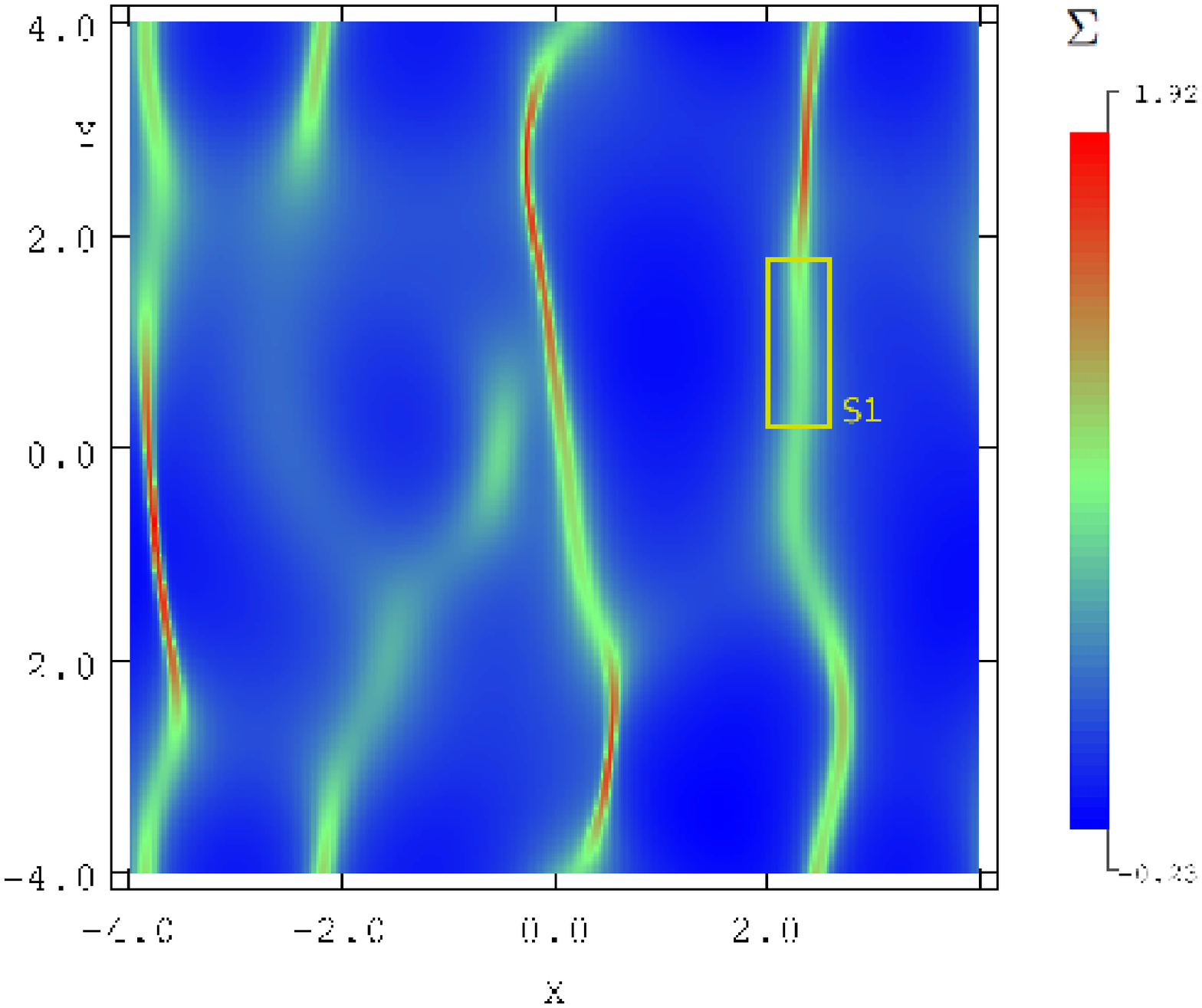}
\caption{Surface density along $z$-axis for the $\beta = 10$ model at $t = 2$ (top) and 
the $\beta = 0.1$, model at $t = 2.25$ (bottom). In terms of the default layer, the area shown is 
8$\times$8 pc$^2$ and the unit of surface density is 5$\times 10^{20}~{\rm cm^{-2}}$.}
\label{fig:pl_bx_noise}
\end{center}
\end{figure}

\section{Parallel magnetic field} \label{sect:magnetic_field} 
\paragraph{Growth rates} 
The analysis of  \citet[][]{Nagaietal1998} shows that the gravitational instability in 
an equilibrium layer is modified by magnetic fields as they stabilize instabilities 
propagating perpendicular to the field (when the external pressure is small). Instabilities along 
the magnetic field are not affected.    The effect of the magnetic field is expressed as a 
function of a dimensionless parameter $\alpha'$ which for our model parameters is given by 
$\alpha' = 2.15/\beta$.  Thus, our models examine a range of $\alpha'$ between 0.215 and 21.5. 
For values as low as $\alpha' = 1.25$ (i.e. $\beta = 2$) the perpendicular perturbations are already 
strongly suppressed (see e.g. Fig.~4 of \citet[][]{Nagaietal1998}).  

Similar to the hydrodynamical simulations we first examine whether our numerical 
model reproduces the expected linear growth rates. We only test this for $\beta = 0.10$
(i.e. $\alpha' = 21.5$). For perturbations along the magnetic field (i.e. along the 
$x$-axis), we find, as expected, linear growth rates that are marginally smaller than in the 
hydrodynamical model (see Fig.~\ref{fig:growth_rates_blz}). The critical wavelength and 
the maximal wavelength are also the same as before. For perpendicular perturbations
along the $y$-axis, the growth rates are significantly reduced.  For $\lambda_{\rm max}$ we 
find $\omega \approx 0.25$ or about 20 times lower than for an identical perturbation 
along the $x$-axis, while we have $\omega \approx 0.85$ when $4\lambda_{\rm max}$. Note that 
the growth rate for $4\lambda_{\rm max}$ is larger than for $\lambda_{\rm max}$ as 
$\lambda_{\rm max}$ is here the value for the fastest growing wave mode along the $x$-axis 
and not the $y$-axis. The critical and maximal wavelength along the $y$-axis are shifted 
towards longer wavelengths, i.e.  $\lambda_{\rm y,max} \approx 2.5\lambda_{\rm max}$ 
and $\lambda_{\rm y, crit} \approx 1.8 \lambda_{\rm crit}$. The shown models for perturbations
perpendicular to the magnetic field then represent models close to the critical
and maximal wavelength. 
  
The simulations confirm the behavior derived from the linear analysis and show that the 
magnetic field introduces a preferred direction for the growth of perturbations, even for 
high values of $\beta\  (> 1)$.

\begin{figure}
\begin{center}
\includegraphics[width=7cm]{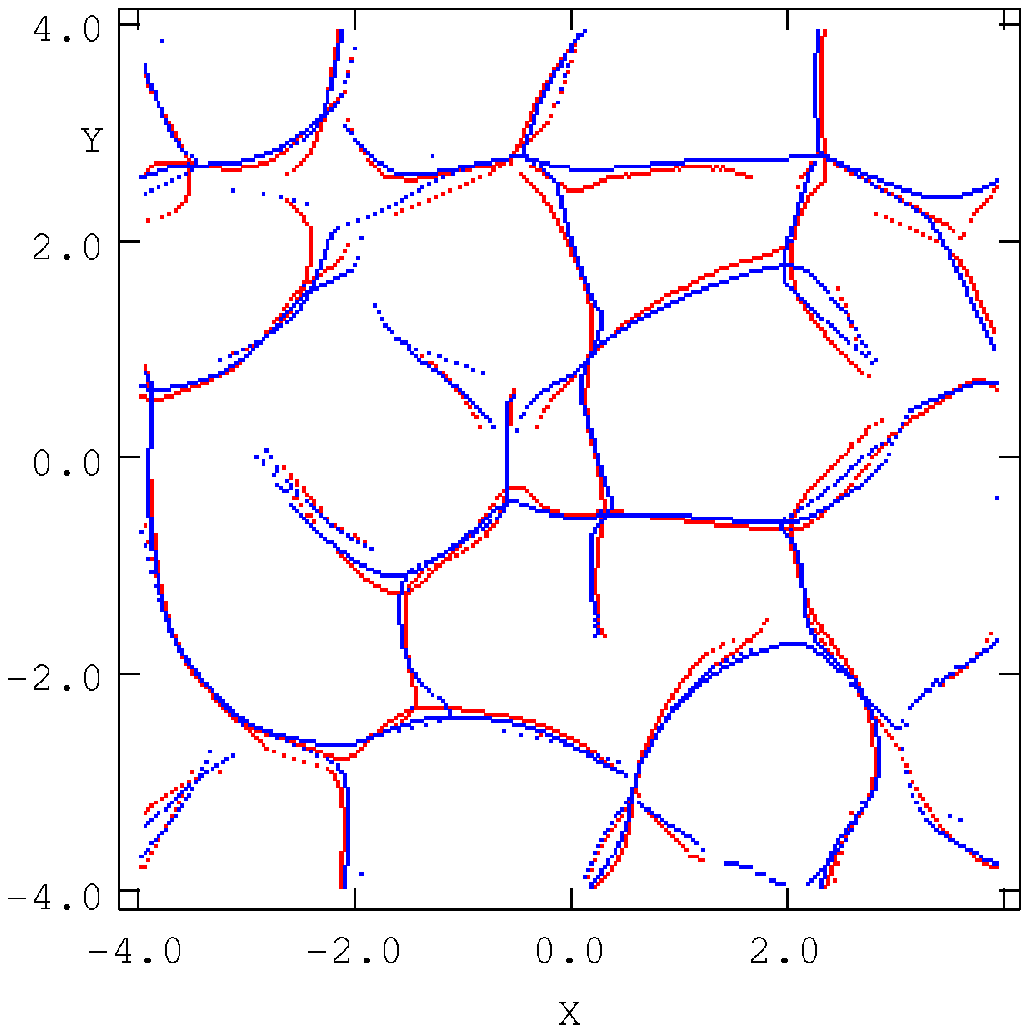}
\includegraphics[width=7cm]{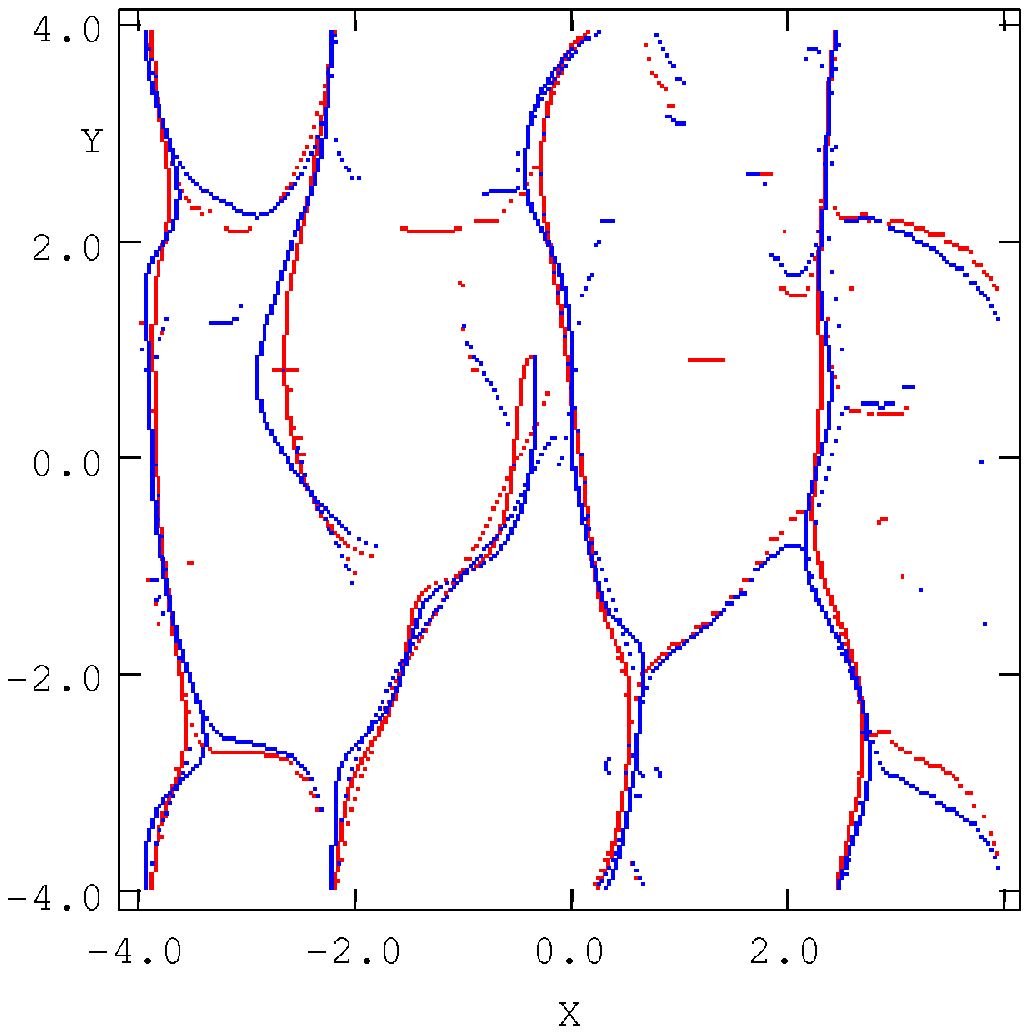}
\caption{Ridges of column density for the weak magnetic field (red) and hydrodynamical (blue) model 
at $t = 2$ (top) and for $\beta = 0.1$ (red) and $\beta = 1$ (blue) at $t = 2.25$ (bottom).}
\label{fig:ridge_pl_bx}
\end{center}
\end{figure}

\paragraph{Filament formation}
We superimpose the equilibrium layer with the same random noise perturbations as the 
hydrodynamical model. Figure~\ref{fig:growth_rate_noise} shows the normalized maximum
perturbation amplitude for $\beta = 10,\ 1,\ {\rm and}\ 0.1$ (as well as for 
$\beta = \infty$, i.e. the hydrodynamical model). The linear phase of the 
instability shows, for all of the magnetic models, the typical growth rate  
of a mode with a wavelength of $\lambda_{\rm max}$. That this specific wave mode
is dominant also becomes apparent from the column density structure (see Fig.~\ref{fig:pl_bx_noise}). 
As  the perpendicular perturbations are significantly damped for the $\beta = 0.1$ 
model (see above), only the modes parallel to the magnetic field become unstable. 
Four parallel filaments are then produced with an separation of about $\lambda_{\rm max}$.
Note that the $\beta = 0.1$ model is shown at $t = 2.25$ and not $t = 2$ as it reaches 
similar densities at later times than in the $\beta = 10$ model.
 
Although the layer is threaded by a magnetic field, the column density structure of 
the $\beta = 10$ model shows filaments both parallel and perpendicular to the magnetic 
field direction.  In fact, the column density structure is very similar to the structure 
seen in the hydrodynamical model. Actually, the positions of the filament ridges are 
identical in both models (see Fig.~\ref{fig:ridge_pl_bx}). The ridge positions for the 
stronger magnetic field models (i.e. $\beta = 1$ and 0.1) are also interchangeable.
Remember that the density initialization is identical for all models and the subsequent
evolution is then determined by the growth of the unstable modes. This means we can 
identify two different regimes, i.e. a quasi-hydrodynamical regime and 
a strong magnetic field one. The separation between the two states lies somewhere between 
$\beta = 2$ and 10. 

\begin{figure}
\begin{center}
\includegraphics[width=8cm]{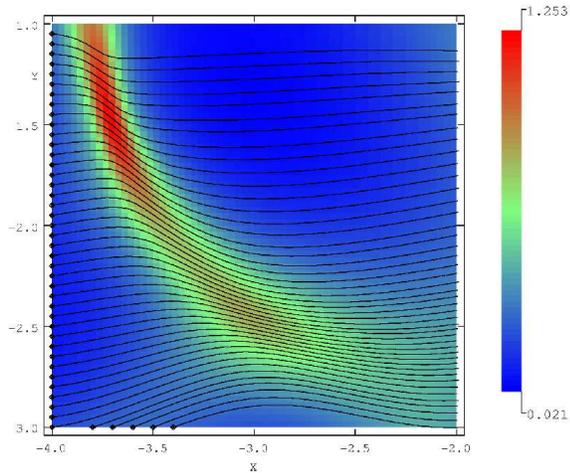}
\caption{The column density of Region~W1 with the vectors showing the mass-weighted 
magnetic field. Their length indicates the magnetic field magnitude with the maximum 
about 3 times the initial magnitude. In terms of the default layer, the unit of column 
density is 5$\times 10^{20}~{\rm cm^{-2}}$ and the maximum magnetic field strength
displayed is 6$\mu$G.}
\label{fig:mag_orientation}
\end{center}
\end{figure}

Not only the density structure, but also the magnetic field structure is distinctively
different in the two regimes. In the strong magnetic regime, the magnetic field is 
dynamically important and suppresses perturbations perpendicular to its orientation.
Then gas flows are predominantly along the magnetic field lines resulting in filaments
perpendicular to the magnetic field. Contrary, in the weak magnetic regime, the magnetic 
field is dynamically passive and is dragged with the gas. Consequently, the density 
structures lie not only perpendicular to the field lines, but also parallel and oblique.

For the oblique filaments an interesting phenomenon occurs: while the magnetic field is
oblique to the filament axis in the outer regions of the filament, it is orientated 
along the filament axis within its central region (see Fig.~\ref{fig:mag_orientation}). 
The filaments are bound by fast-mode shocks in which the magnetic 
field increases perpendicular to the shock normal. The formation of such filaments 
is described in \citet[][]{NakajimaHanawa1996} who show that such filaments 
are in quasi-static equilibrium in their inner parts.

\begin{figure}
\begin{center}
\includegraphics[width=8cm]{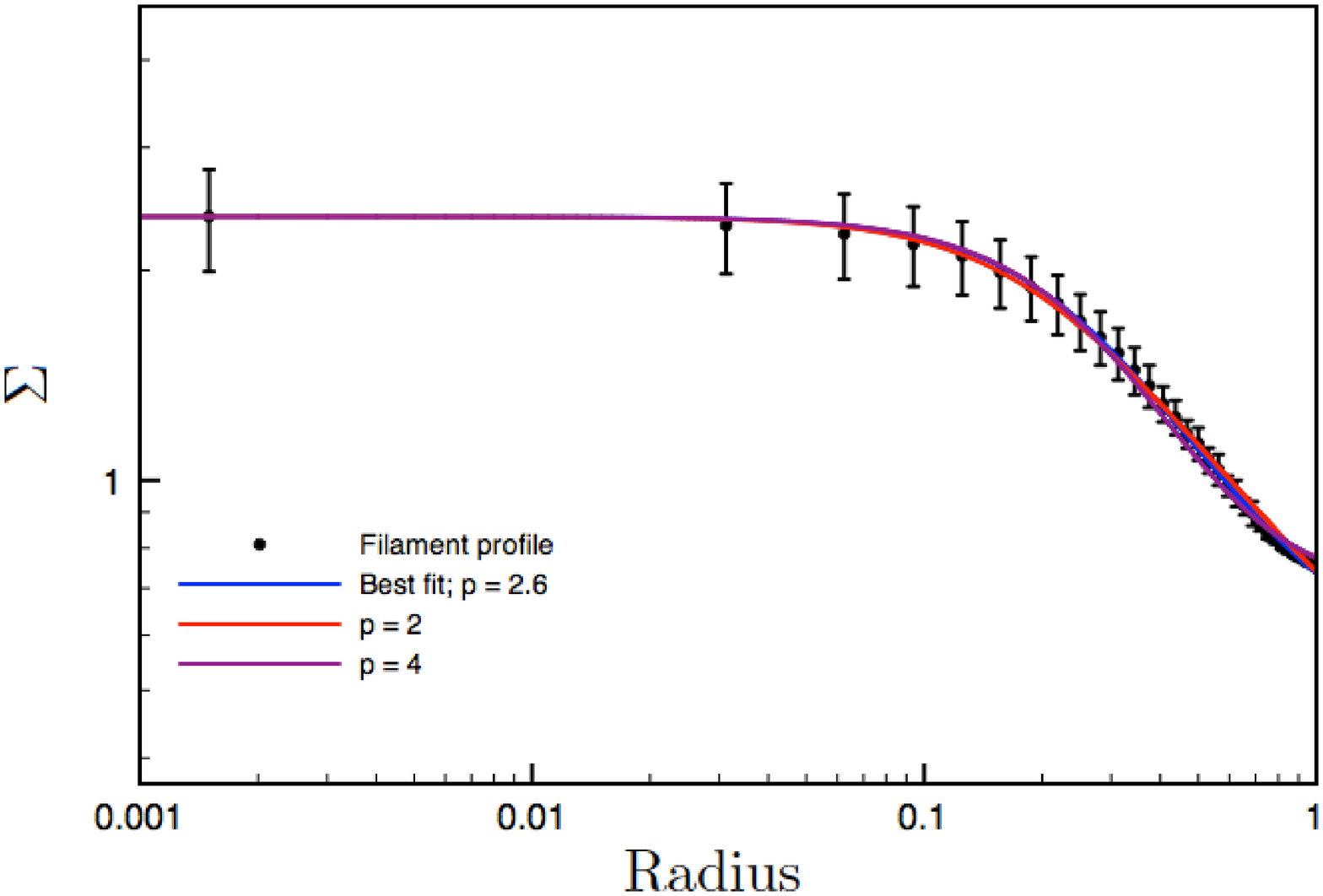}
\includegraphics[width=8cm]{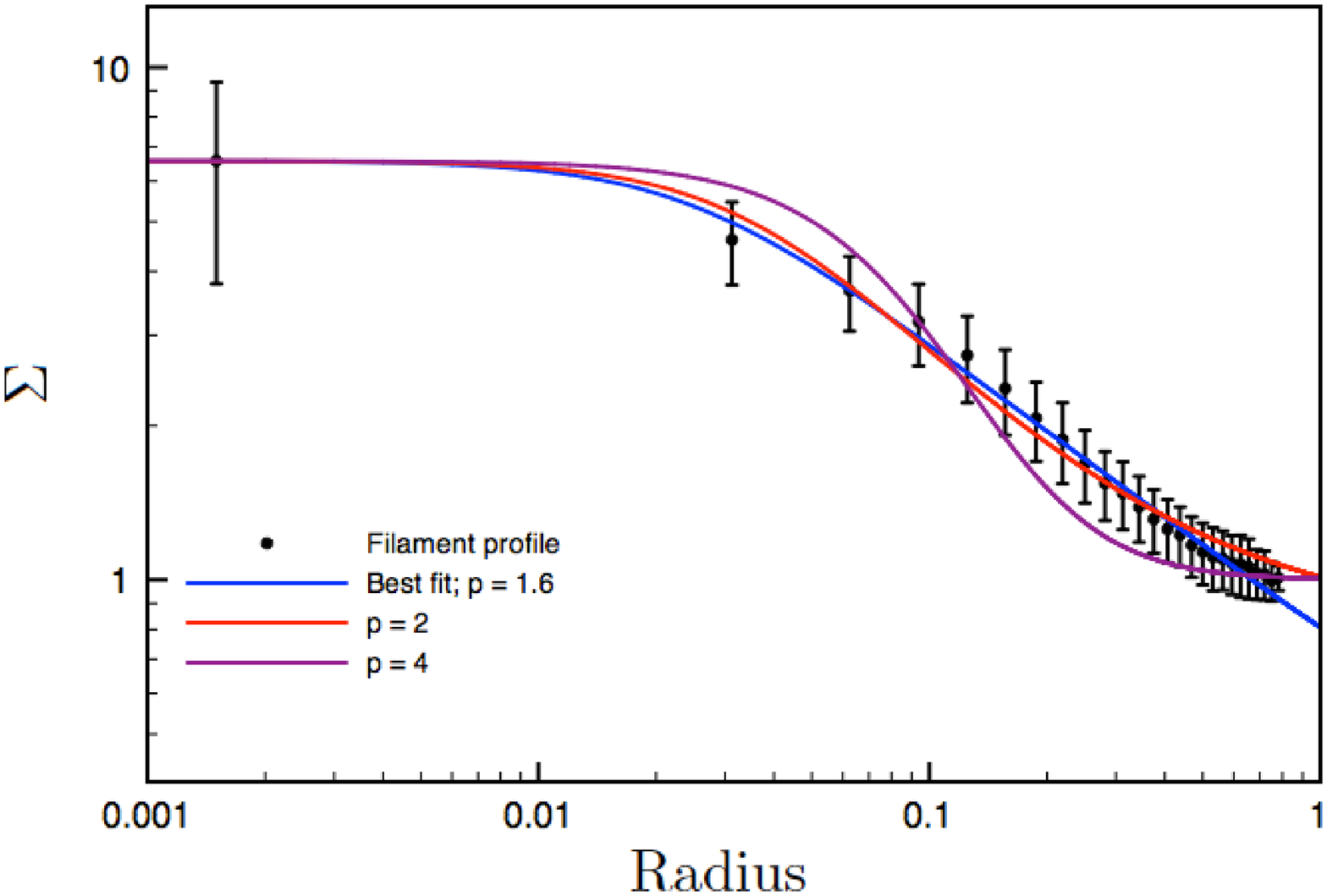}
\caption{Radial column density profiles for Region~W1 (top) and Region~W2 (bottom)
averaged over the entire filament. The error bars show the dispersion on the mean value
at a given radius. The solid lines show the best fit (blue), a $p = 2$ (red) and $ p = 4$ 
(purple) Plummer profile. For the default layer, the units of column density and length are
5$\times 10^{20}~{\rm cm^{-2}}$ and 1~pc, respectively.}
\label{fig:pl_bhx_profile}
\end{center}
\end{figure}

\begin{figure}
\begin{center}
\includegraphics[width=8cm]{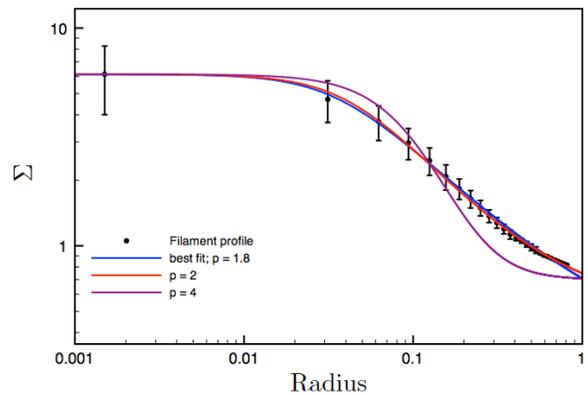}
\caption{Average radial column density profile for Region~S1 with the dispersion on the mean
given by the error bars.  The solid lines show the best fit (blue), a $p = 2$ (red) and 
$ p = 4$ (purple) Plummer profile. For the default layer, the units of column density and length 
are 5$\times 10^{20}~{\rm cm^{-2}}$ and 1~pc, respectively. }
\label{fig:pl_blx_profile}
\end{center}
\end{figure}

\begin{figure}
\begin{center}
\includegraphics[width=8cm]{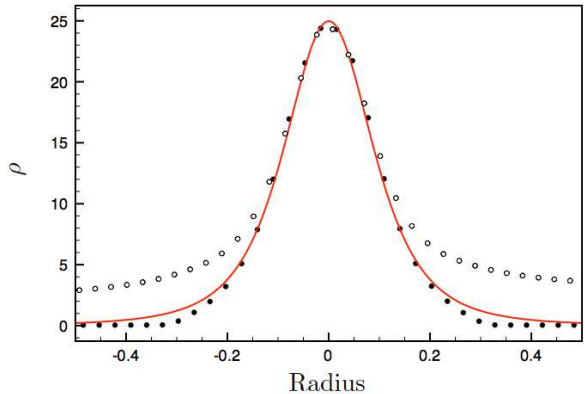}
\caption{Density slice at $y = 1$ for Region~S1. The solid circles show the density along the 
$z$-axis and the open circles along the $x$-axis. The solid line is the profile for
an Ostriker cylinder with $\rho_c = 24.9$. In terms of the default layer, the length unit
is 1~pc and the density unit 159.2 cm$^{-3}$.}
\label{fig:pl_blx_slice}
\end{center}
\end{figure}

\paragraph{Radial filament profile}
The average radial column density profile for the filaments in the magnetized layers 
is similar to the profiles extracted for the hydrodynamical model (see 
Figs.~\ref{fig:pl_bhx_profile} and \ref{fig:pl_blx_profile}). The distribution is 
best fitted by Plummer profiles (Eq.~\ref{eq:plummer}) with power-law indices $p \leq 2$
and cannot be fitted by an equilibrium profile with $p = 4$. An exception is 
Region~W1 which can be fitted reasonably well by a large range of $p$-values, including
$p = 4$. 
The ratio of the central-to-external column density for Region~W1 is small.
The distribution then does not sample the power-law tail of the Plummer profile properly as 
the distribution is dominated by the flat inner region and the background
column density.  This results in a broad range of possible power-law indices for the 
Plummer profile.

Again, density slices through the filaments show that the density distribution of the central
region can be approximated by an equilibrium Schmid-Burgk profile. For example,  the density
distribution for a slice at $x = 1.25$ in Region~W1 is best fitted with $A = 0.13$ and 
$\rho_c = 4.825$. A slice through $y = 1$ of Region~S1 is best fitted by an Ostriker cylinder 
(or a Schmid-Burgk profile with $A = 1$) with $\rho_c = 24.9$ (see Fig.~\ref{fig:pl_blx_slice}).  
The selection of these filaments represents different stages during the formation
process, i.e. the filament of Region~W1 is starting to form, while Region~S1 shows the end of the 
formation process.
Although the interpretation from the density profiles
seems inconsistent with the one inferred from the average column 
density distribution, it is not.  For filaments exceeding the critical line-mass, the density 
distribution is described by an equilibrium profile only for radii below the scale height. 
The excess mass is at radii above the scale height resulting in a radial profile flatter 
than $r^{-4}$. As the Plummer profile only measures the radial dependence at radii 
above the scale height, such a description misses the equilibrium distribution in the 
center of the filament.

A more important result is that the filaments can be approximated by a hydrodynamical equilibrium 
without adjusting the sound speed (to take into account additional support provided by magnetic 
support).  This suggests that the magnetic field is not important in setting the density 
distribution inside of the filaments.  In filaments perpendicular to the magnetic field, gas flows 
along the field lines.  Then magnetic pressure gradients are not generated and are thus much smaller 
than pressure (or density) gradients. In the quasi-hydrodynamical regime, filaments
also form parallel to the magnetic fields. Because of magnetic flux conservation,
a scaling of the magnetic field with density is established, i.e. $B \propto \rho$.
The magnetic pressure, then, has a similar slope as the thermal pressure.
However, as $\beta > 1$, the magnetic pressure gradient is still much smaller
than the thermal pressure gradient and does not contribute to the total pressure
force balancing self-gravity.

\paragraph{Dense core formation} The formation of dense cores is different in the 
two regimes. While cores form at the junctions of filaments in the weak magnetic regime,
the cores in the strong magnetic regime form due to fragmentation of the filament. However,
\citet[][]{InutsukaMiyama1992} show that the filaments only fragments if the growth rate of 
the unstable axisymmetric is shorter than the radial collapse time. This condition is 
only fulfilled if $m_{\rm line} \approx m_{\rm crit}$. For our initial conditions,
the line mass is much larger than the critical value, i.e. $m_{\rm line} = 1.9\ m_{\rm crit}$,
and, thus,  we do not expect cores to form within the filaments due to fragmentation. The 
condensations that we see in the filaments (see Fig.~\ref{fig:pl_bx_noise}) are actually a 
result of the initial conditions that break the uniformity of the layer.

\begin{figure}
\begin{center}
\includegraphics[width=8cm]{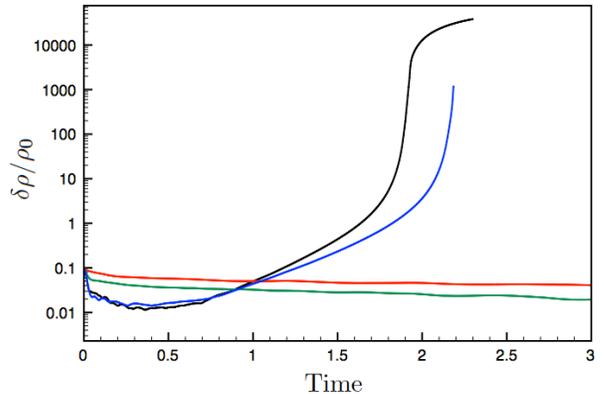}
\caption{Similar figure as Fig.~\ref{fig:growth_rates_hydro} but for perpendicular magnetic fields.
The solid lines represent a hydrodynamical (black), $\beta = 10$ (blue), 1 (red) and 0.1 (green) 
model. The unit of time for our default layer is 4.9Myr.}
\label{fig:growth_rates_bz}
\end{center}
\end{figure}

\section{Perpendicular magnetic field} \label{sect:perp_magnetic_field} 
The previous section discusses the gravitational instability in a layer threaded by 
a parallel magnetic field. Here, we examine the fragmentation for a magnetic field perpendicular
to the layer. \citet[][]{NakanoNakamura1978} study the marginally stable
modes of an isothermal layer with perpendicular magnetic field. They find that the layer is 
unstable to gravitational perturbations when $\Sigma_0/B_0 > (\pi G)^{-1/2}$ or simply $\beta > 1$,
with the critical wavelength close to the hydrodynamical value. This means that, contrary 
to a parallel magnetic field, a perpendicular magnetic field can stabilize the layer as long 
as it is strong enough. Figure~\ref{fig:growth_rates_bz} indeed shows that perturbations in 
models with $\beta \leq 1$ are damped. For these layers to become unstable, the magnetic
support needs to disappear by e.g. ambipolar diffusion \citep[][]{Kudohetal2007, KudohBasu2008}. 

\begin{figure}
\begin{center}
\includegraphics[width=8cm]{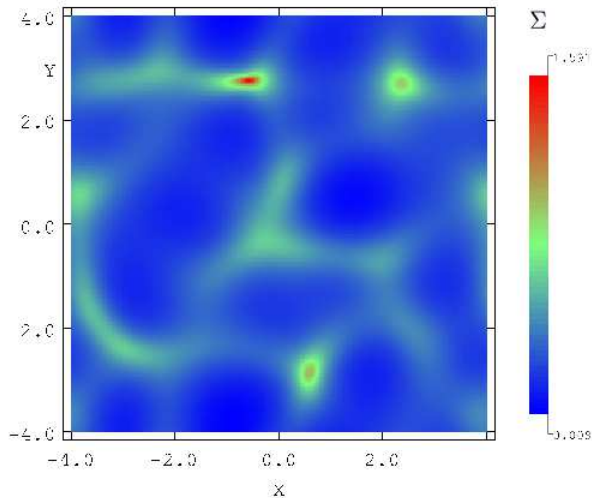}
\caption{Column density structure along the $z$-axis at $t = 2.15$ for a perpendicular magnetic 
field with $\beta = 10$. For the default layer, the unit of column density is 
5$\times 10^{20}~{\rm cm^{-2}}$.}
\label{fig:pl_bhz_noise}
\end{center}
\end{figure}

\begin{figure}
\begin{center}
\includegraphics[width=8cm]{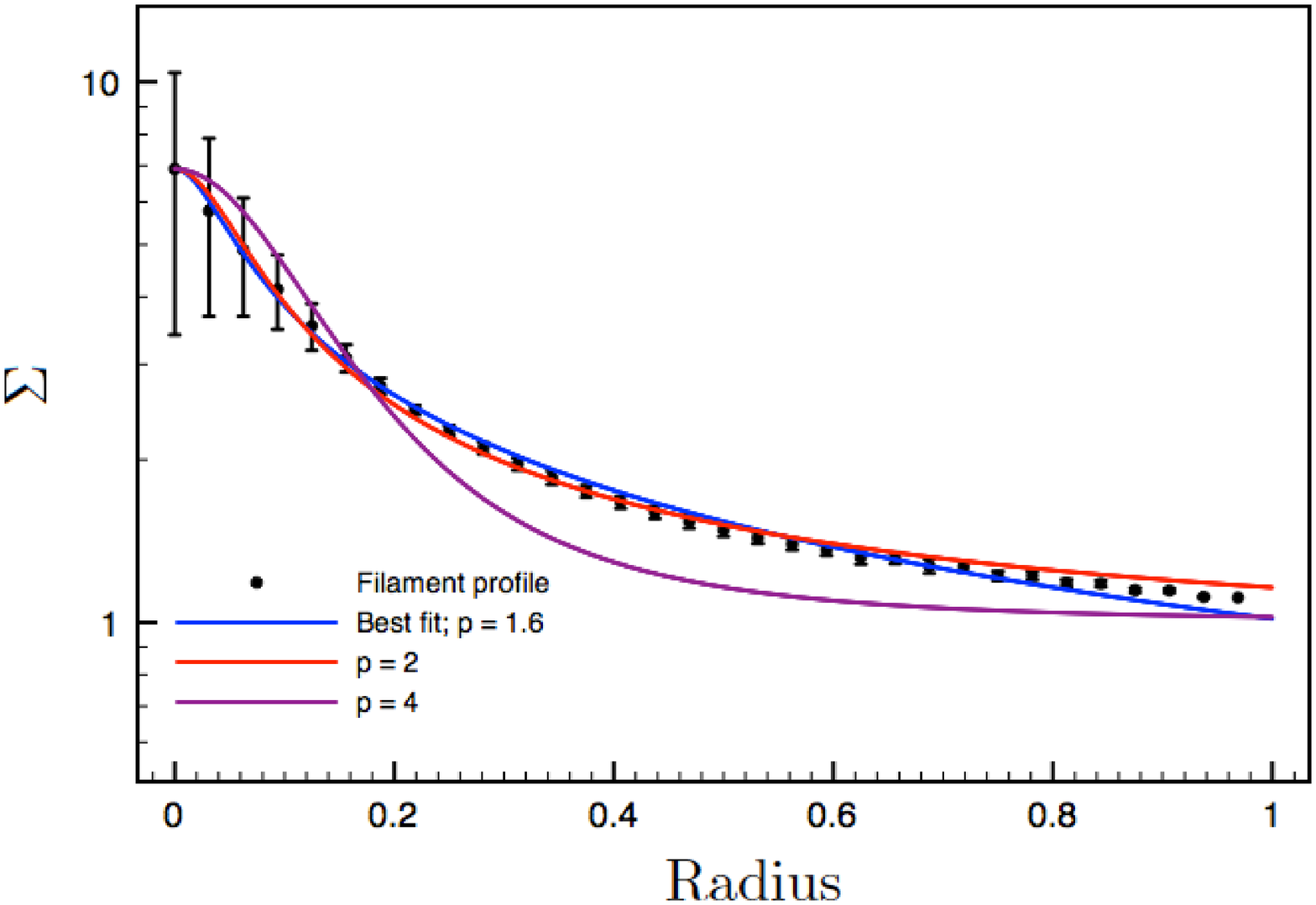}
\includegraphics[width=8cm]{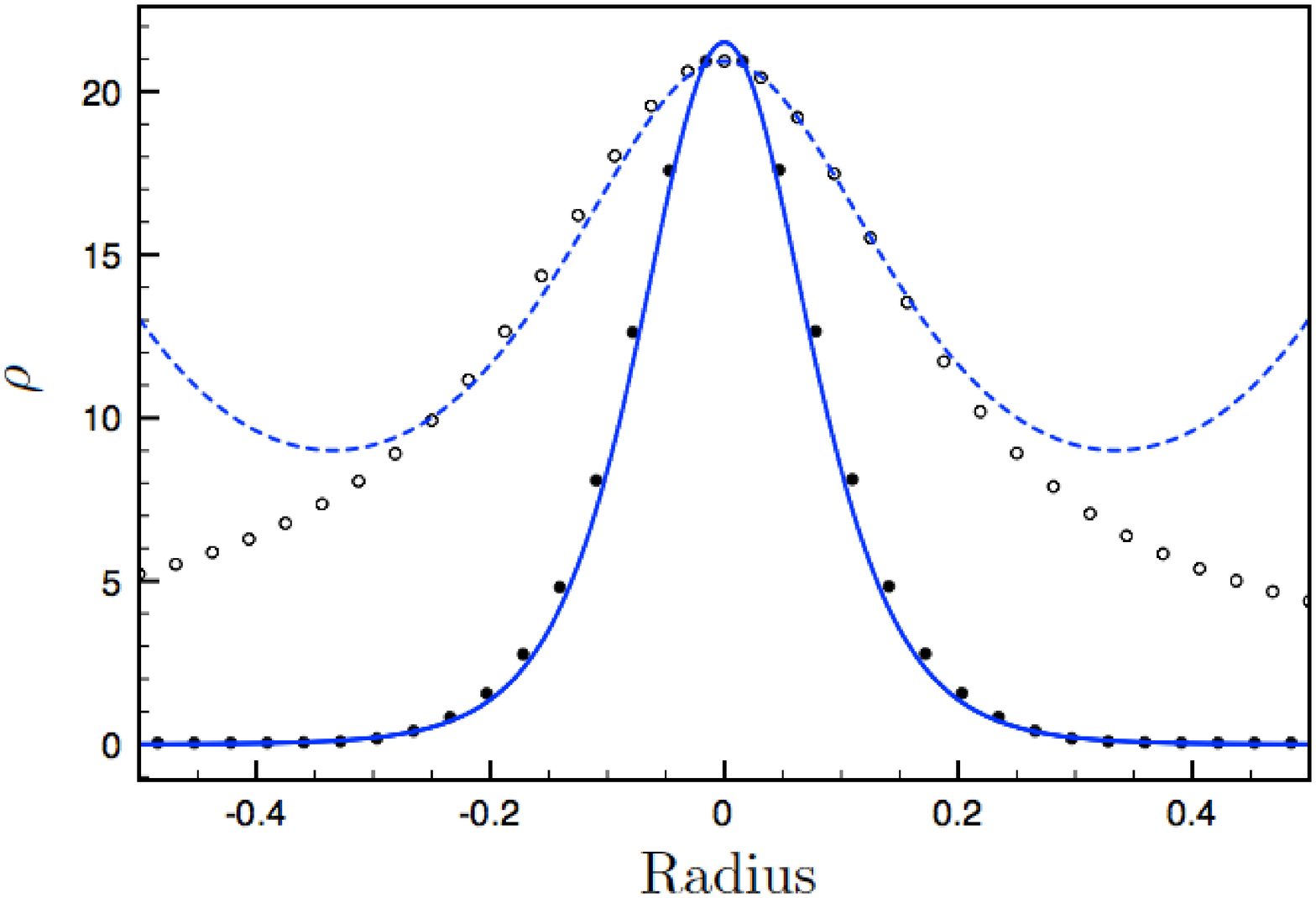}
\caption{{\it Top:} Average column density profile of a selected filament in Fig.~\ref{fig:pl_bhz_noise}.
The error bars represent the dispersion on the average value at a given radius.
The solid lines are the best-fit (blue), a $p = 2$ and $p = 4$ Plummer profile. The unit of 
column density of the default layer is 5$\times 10^{20}~{\rm cm^{-2}}$ and the unit of length is 1~pc.
{\it Bottom:} Density profile along the $x$ (open circles) and $z$-axis (solid circles) for a slice 
through the filament. The blue lines show the Schmid-Burgk profile with $A = 0.29$ and $\rho_c = 11.81$
along the $x$ (dashed) and $z$ (solid) axis. For the default layer, the unit of length is 159.2~cm$^{-3}$.}
\label{fig:pl_bhz_prof}
\end{center}
\end{figure}

\paragraph{Density structure and filaments} The $\beta = 10$ model has a similar, though 
slightly slower, growth rate as the hydrodynamical model. This reduced growth rate is due to 
magnetic tension forces acting to straighten field lines (which are bended because
the flow in the layer is perpendicular to the field lines). The mode associated with 
this growth still has a wavelength close to $\lambda_{\rm max} = 4\pi H$, as the 
resulting column density structure is very similar to the hydrodynamical model 
(see Fig.~\ref{fig:pl_bhz_noise}). The filament ridges are actually in the same places.

Not only does the column density structure look very similar, the column density and 
density profiles of the filaments show the same properties as in the previous models 
(see Fig.~\ref{fig:pl_bhz_prof}). The power index of the best-fit Plummer 
profile has a value of $\leq 2$, but, at the same time, the central density structure 
of the filaments is perfectly described by an equilibrium
Schmid-Burgk profile. Again, this means that the magnetic field only contributes marginally
to the dynamics. The plasma $\beta$ near the center of the filament decreases but is still 
above $\beta = 1$.    

\begin{figure}
\begin{center}
\includegraphics[width=7cm]{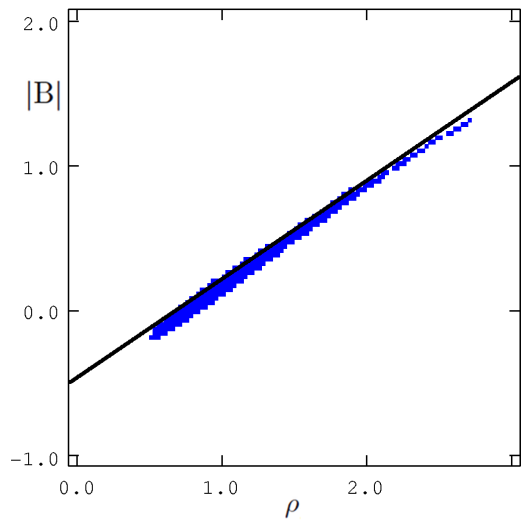}
\caption{Log-log relation between magnetic field magnitude and density at the midplane for 
the perpendicular $\beta = 10$ model. The solid line shows the dependence for an isotropically
contracting sphere. In terms of the default layer, the units of the magnetic field and density
are given by 1.76$\mu$G and 152.9~cm$^{-3}$.} 
\label{fig:Bdens}
\end{center}
\end{figure}

\paragraph{Dense cores} While the magnetic field does not influence filaments to any significant 
degree, the evolution of the cores are affected by them. Dense cores still arise at the junctions 
of the filaments, but the density distribution of the cores is very much axisymmetric.
In the hydrodynamical model, gas is predominantly fed to the cores along the filaments and not 
isotropic as suggested here. Because of magnetic tension forces, gas in the vicinity of the cores 
flows directly towards the core instead of first to a filament and then towards the core. 
The former produces a magnetic field that is less `twisted'.
The axisymmetry of the dense core is also observed in the scaling relation of the magnetic field
and density relation, i.e. $|B| \propto \rho^{2/3}$ (see Fig.~\ref{fig:Bdens}). Such a relation is 
obtained for an isotropically contracting sphere threaded by a frozen-in magnetic field.
Although the magnetic energy increases more rapidly than the thermal pressure in the 
core (i.e. $\beta \propto \rho^{-1/3}$), the magnetic field is never strong enough to 
prevent collapse. The potential energy of a spherical core increases at the same rate as the 
magnetic energy during contraction. Initially, the gravitational energy is larger than the 
magnetic energy and thus remains so during the subsequent evolution.

\begin{figure*}
\begin{center}
\includegraphics[width=16cm]{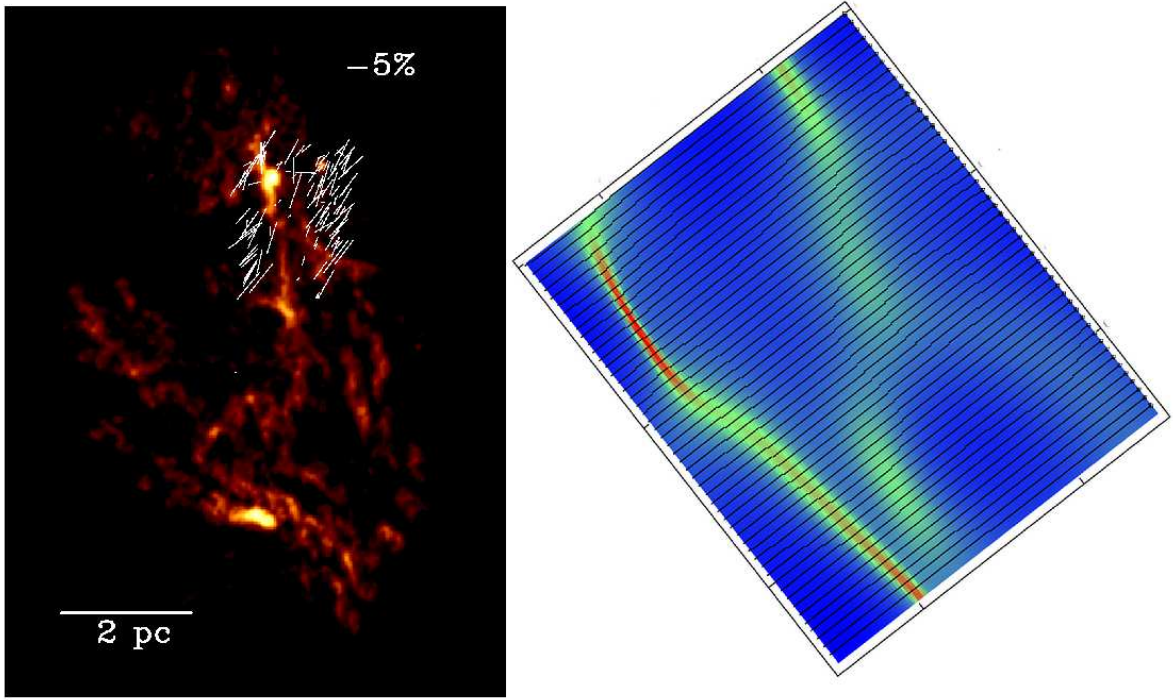}
\caption{{\it Left:} NH$_3$ emission map of the massive star forming complex G14.225-0.506
\citep{Busquetetal2013}. The white lines show the polarization vectors of the near-infrared 
(H-band) polarimetric observations \citep{Busquetetal2013}. {\it Right:} Section of 
the bottom panel of Fig.~\ref{fig:pl_bx_noise}. The solid lines represent the mass-weighted 
magnetic field integrated along the $z$-axis.}
\label{fig:G14}
\end{center}
\end{figure*}

\section{Comparison with the star-forming region G14.225-0.506}\label{sect:comparison}
Recently, \citet{Busquetetal2013} reported VLA observations in 
NH$_3$ of a massive star forming complex G14.225-0.506. The NH$_3$ emission reveals 
a network of filaments that are aligned in parallel in projection on to the plane 
of the sky (see Fig.~\ref{fig:G14}). The filaments appear to take two directions, 
one group is oriented at a position angle (PA) of 10$^\circ$, and the other is at a PA of 
60$^\circ$. The averaged projected separation between the adjacent filaments are between 
0.5 pc and 1 pc. Polarimetric observations in the near infrared H-band toward the 
northern section of the complex reveal magnetic field orientations perpendicular to the main 
axis of the filaments. 

The fragmentation of a layer threaded with a strong parallel magnetic field reproduces  
the G14.225-0.506 filament network (see Fig.~\ref{fig:G14}). This allows to 
infer the initial conditions of the initial, unperturbed layer. The separation of 
the adjacent filaments is given by $\lambda_{\rm max} = 4\pi H$ which results
in a scale height of 0.04-0.08~pc for the unperturbed layer. Then the initial surface 
density is $\Sigma = a^2/(\pi G H) \approx 1.5-3\times 10^{-2} {\rm g\ cm^{-2}}$ (assuming 
a sound speed of 0.2 ${\rm km\ s^{-1}}$). As the area covered by the filaments is 
4.7$\times$ 8.7 pc$^2$, the unpertubed layer contains about 2900-5800~M$_\sun$. 
This agrees well with the combined mass of 2600~M$_\sun$ for the filaments in G14.225-0.506 
\citep{Busquetetal2013}. Also, an estimate of the magnetic field strength can 
be made. For the strong parallel field model, $\beta < 1$ and the required magnetic field 
is stronger than 12-25 $\mu$G. Zeeman measurements of the magnetic field indicate that 
these values are towards the high end of, but within, the observed range in magnetic field strengths
\citep[][]{Crutcher2012}. 

The initial conditions needed to produce the density structure of the filamentary 
complex G14.225-0.506 by gravitational instability in the presence of a strong magnetic field
 are not unrealistic. However, we cannot exclude any other formation process such as e.g. colliding flows.

\section{Discussion and conclusions}\label{sect:discussion_and_conclusions}
In this paper we have examined the gravitational instability in an isothermal equilibrium layer
threaded by a magnetic field. Our main results are as follows.

\begin{enumerate}
\item An equilibrium isothermal layer threaded by parallel magnetic fields is 
always unstable for gravitational instabilities with wavelengths larger 
than $\lambda_{\rm crit} = 2\pi H$. For perpendicular magnetic fields instabilities 
only arise when $\beta > 1$. The density structures of the unstable layer can 
be described by two different regimes, i.e. a hydrodynamical one 
and a strong magnetic one. The network structure of filaments can then be used to
estimate the magnitude of the magnetic field. Parallel filaments indicate that the magnetic 
field is strong with $\beta \leq 1$ and that the field lines lie within the plane of 
the layer.  On the other hand, a filament-hub structure implies that the magnetic field 
is weak with $\beta \gg 1$. In this case there is no connection between 
filament axes and the magnetic field, i.e. the magnetic field can lie parallel or 
perpendicular along the filament axes.

\item The filaments that form have a line-mass that exceeds the critical value for 
axisymmetric collapse  and thus cannot be in hydrostatic equilibrium.
However, at any time during their evolution, the
central regions of the filaments are described by an equilibrium density 
structure.  Using a Schmid-Burgk (1967) density profile (Eq.~\ref{eq:Schmid}), the evolution 
of a filament from a perturbation in the equilibrium layer ($A \approx 0$) towards an 
Ostriker (1964) distribution  ($A = 1$) is perfectly captured. 
Furthermore, because of the thermal support
in the central region of the filament, the collapse time is much longer than the 
free-fall time scale (similar to the results of \citet[][]{InutsukaMiyama1992} for 
collapsing cylinders).

Although the central region of filaments is well described by equilibrium profiles,
the distribution for radii above the scale height deviates from the equilibrium (as 
that region contains an excess of mass). The profile is then flatter than the expected
power law of $r^{-4}$ and similar to the results from {\it Herschel} observations
\citep[e.g.][]{Arzoumanianetal2011}. An additional flattening of the profile is obtained by 
averaging the column density profile along a filament with a central density 
variation. An increase in the central density produces a larger central column density
but also implies a smaller scale height (i.e. $R_0 \propto \rho^{-1/2}$). 
Such a variation in the filament width is actually observed along the L1506 filament
in the Taurus molecular complex \citep[][]{Ysardetal2013}. 

By studying colliding flows, \citet[][]{Gomez&Vazquez2014} show that filaments 
also form due to dynamical and thermal instabilities in an initially sheet-like structure.
While their filaments exhibit the flat column density profile of the observations and 
our simulations, their formation and evolution does not follow quasi-equilibrium states. 
\citet[][]{Gomez&Vazquez2014} argue that the filaments are just long-lived, persistent
features of the flow. However, the resolution of their simulation is not high enough
to resolve the thermal scale height associated with the filaments (e.g. H $\approx$ 0.3 pc 
for their filament 1, while the resolution is $\approx 0.17$ pc). Without the proper resolution,
thermal pressure forces cannot balance gravity numerically. We encounter the same problem 
when we analyze filaments with a too high central density ($\rho_c > 150$ in code units). 
We also recognize the difficulty in resolving the thermal scale height in turbulent 
simulations where the density contrasts fluctuate rapidly.

\item Our simulations do not reproduce the other property of filaments, i.e. 
a near constant FWHM of $\sim 0.1$ pc. As the central density of the filaments 
increases, the scale height continues to decrease. As the scale height also 
depends on pressure support, i.e. $R_0 \propto a$, the pressure support within the 
filament needs to increase as it is collapsing. Our simulations show that the magnetic
fields are not able to provide this extra support. 

However, \citet[][]{KlessenHennebelle2010}
show that turbulence can be driven by accretion.  Interstellar filaments indeed 
show that the filament velocity dispersion increases with column density 
and exceeds the thermal sound speed if they have a line-mass above the critical value
\citep{Arzoumanianetal2013}.

\item 
Different authors \citep[e.g.][]{Li_etal2004, Banerjee_etal2009, Collins_etal2011} show
that the magnetic field does not prevent the formation and collapse of dense structures
in molecular clouds. However, these studies only consider a low level of magnetization.
In case of a strong magnetic field, simulations show that the field dominates turbulence 
and inhibits the gravitational collapse \citep[e.g.][]{Nakamura&Li2008, Basu&Dapp2010}.
The filaments in our simulations are purely hydrodynamical structures. This means that the 
magnetic field does not play a significant role in setting the shape of the filaments, 
even in a low-beta plasma. 
 
\item Dense cores form at the same time as the filaments. In the hydrodynamical 
regime, the cores arise at the junctions of the filaments. Mass accretion of the cores 
depends on the orientation of the magnetic field, i.e. for perpendicular fields the accretion is 
roughly isotropic, while, for parallel fields, much of the gas is fed along the filaments 
and thus highly directional. Such non-isotropic mass accretion drastically changes the observational
signatures of dense core collapse \citep[e.g.][]{Smithetal2012, Smithetal2013}. Enough mass is  
accreted for the cores to contract, but at rates slower than free-fall collapse. This again 
suggests that thermal pressure (and magnetic 
pressure to a lesser degree) are able to roughly balance against self-gravity. 
\end{enumerate}

Some caveats need to be taken into account when discussing these findings.
Turbulent motions are ubiquitous in the ISM and it is unlikely that a thick, quiescent 
equilibrium layers forms out of such a medium. For example, simulations of colliding flows 
\citep[e.g.][]{Vazquezetal2003, AuditHennebelle2005, Heitschetal2008} show that the 
shock-bounded layer is prone to dynamical instabilities, such as Kelvin-Helmholtz and Raleigh-Taylor
instabilities, but also to thermal instabilities. However, \citet[][]{KudohBasu2011} show that, 
dense core profiles do not not depend strongly on the initial velocity perturbations, although
their formation time scales are shorter for larger velocity fluctuations. 

For a thin isothermal layer, the nature of the gravitational instability changes and is 
very similar to an incompressible mode. \citet[][]{Nagaietal1998} show that the density perturbations 
are then a result of a deformation of the boundary of the layer. When the layer is threaded 
by a strong magnetic field ($\beta < 1$), filaments no longer form perpendicular to the 
magnetic field lines but parallel to them. Furthermore, the line-mass of the filaments 
is smaller than the critical value. Then the filaments do not longer collapse but are in 
equilibrium. It is then expected that the filament fragments along its axis into cores
with equal spacing between them. We will study this possibility in a subsequent paper.

\acknowledgements We thank Tom Hartquist, Sam Falle and Phil Myers for useful discussions,
Gemma Busquet for providing the observational data of the G14.225-0.506 molecular cloud
and the anonymous referee for his/her comments that improved the paper.
SvL acknowledges support from the SMA Postdoctoral Fellowship of the Smithsonian Astrophysical Observatory.
The simulations for this work were run on the {\it Smithsonian Institution High Performance 
Cluster} (SI/HPC).


\begin{thebibliography}{}
\bibitem[Alves et al.(2001)]{Alvesetal2001}
     Alves, J.~F., Lada, C.~J., \& Lada, E.~A.\ 2001, \nat, 409, 159
\bibitem[Andr\'e et al.(2010)]{Andreetal2010}
     Andr\'e, P., Men'shchikov, A., Bontemps, et al. 2010, A\&A, 518, L102+
\bibitem[Arzoumanian et al.(2011)]{Arzoumanianetal2011}
     Arzoumanian, D., Andr\'e, P., Didelon, P., et al. 2011, A\&A, 529, L6+
\bibitem[Arzoumanian et al.(2013)]{Arzoumanianetal2013}
     Arzoumanian, D., Andr{\'e}, P., Peretto, N., Konyves, V.\ 2013, \aap, 553, A119
\bibitem[Audit \& Hennebelle(2005)]{AuditHennebelle2005}
     Audit, E., \& Hennebelle, P.\ 2005, \aap, 433, 1
\bibitem[Ballesteros-Paredes et al.(1999)]{Ballesterosetal1999}
     Ballesteros-Paredes, J., Hartmann, L., \& V\'azquez-Semadeni, E. 1999,
     \apj, 527, 285
\bibitem[Ballesteros-Paredes \& Mac Low(2002)]{BallesterosMacLow2002}
     Ballesteros-Paredes, J., \& Mac Low, M.-M.\ 2002, \apj, 570, 734
\bibitem[Ballesteros-Paredes et al.(2007)]{Ballesterosetal2007}
     Ballesteros-Paredes, J., Klessen, R.~S., Mac Low, M.-M.,
     \& Vazquez-Semadeni, E.\ 2007, Protostars and Planets V, 63
\bibitem[Banerjee et al.(2009)]{Banerjee_etal2009} Banerjee, R., V{\'a}zquez-Semadeni, 
     E., Hennebelle, P., \& Klessen, R.~S.\ 2009, \mnras, 398, 1082 
\bibitem[Basu \& Dapp(2010)]{Basu&Dapp2010} Basu, S., \& Dapp, W.~B.\ 2010, \apj, 716, 427 
\bibitem[Bonnor(1956)]{Bonnor1956}
     Bonnor, W.~B.\ 1956, \mnras, 116, 351
\bibitem[Busquet et al.(2013)]{Busquetetal2013}
     Busquet, G., Zhang, Q., Palau, A., et al.\ 2013, \apjl, 764, L26
\bibitem[Collins et al.(2011)]{Collins_etal2011} Collins, D.~C., Padoan, P., Norman, 
     M.~L., \& Xu, H.\ 2011, \apj, 731, 59
\bibitem[Crutcher(2012)]{Crutcher2012} Crutcher, R.~M.\ 2012, \araa, 50, 29
\bibitem[Curry(2000)]{Curry2000}
     Curry, C.~L.\ 2000, \apj, 541, 831
\bibitem[de Avillez \& Breitschwerdt(2005)]{deAvillezBreitschwerdt2005}
     de Avillez, M.~A., \& Breitschwerdt, D.\ 2005, \aap, 436, 585
\bibitem[Dedner et al.(2002)]{Dedneretal2002}
     Dedner, A., Kemm, F., Kr{\"o}ner, D., et al.\ 2002, Journal of
     Computational Physics, 175, 645
\bibitem[Elmegreen \& Elmegreen(1978)]{ElmegreenElmegreen1978}
     Elmegreen, B.~G., \& Elmegreen, D.~M.\ 1978, \apj, 220, 1051
\bibitem[Falle(1991)]{Falle1991}
     Falle S.~A.~E.~G., 1991, MNRAS, 250, 581
\bibitem[Falle et al.(2012)]{Falleetal2012}
     Falle, S., Hubber, D., Goodwin, S., \& Boley, A.\ 2012, Numerical Modeling
     of Space Plasma Slows (ASTRONUM 2011), 459, 298
\bibitem[Fiege \& Pudritz(2000)]{FiegePudritz2000}
     Fiege, J.~D., \& Pudritz, R.~E.\ 2000, \mnras, 311, 105
\bibitem[Fischera \& Martin(2012)]{FischeraMartin2012}
     Fischera, J., \& Martin, P.~G.\ 2012, \aap, 542, A77
\bibitem[Galv{\'a}n-Madrid et al.(2013)]{GalvanMadridetal2013} Galv{\'a}n-Madrid, R.,
     Liu, H.~B., Zhang, Z.-Y., et al.\ 2013, \apj, 779, 121
\bibitem[Goldsmith et al.(2008)]{Goldsmithetal2008}
     Goldsmith, P., Heyer, M., Narayanan, G., Snell, R., Li, D.,
     \& Brunt, C. 2008, ApJ, 680, 428
\bibitem[Gomez \& Vazquez-Semadeni(2013)]{Gomez&Vazquez2014} Gomez, G.~C., \& 
     Vazquez-Semadeni, E.\ 2013, arXiv:1308.6298 


\bibitem[Goodman et al.(1990)]{Goodmanetal1990}
     Goodman, A. A., Bastien, P., Myers, P. C., \& M\'enard, F. 1990,
     \apj, 359, 363
\bibitem[Goodman et al.(1992)]{Goodmanetal1992}
     Goodman, A. A., Jones, J. T., Lada, E. A., \& Myers, P. C. 1992,
     \apj, 399, 108
\bibitem[Heitsch(2013)]{Heitsch2013}
     Heitsch, F.\ 2013, \apj, 769, 115
\bibitem[Heitsch et al.(2008)]{Heitschetal2008}
     Heitsch, F., Hartmann, L.~W., \& Burkert, A.\ 2008, \apj, 683, 786
\bibitem[Inutsuka \& Miyama(1992)]{InutsukaMiyama1992}
     Inutsuka, S., \& Miyama, S. M. 1992, \apj, 388, 392
\bibitem[Johnstone \& Bally(1999)]{JohnstoneBally1999}
     Johnstone, D., \& Bally, J. 1999, ApJ, 510, L49
\bibitem[Keto et al.(2004)]{Ketoetal2004}
     Keto, E., Rybicki, G.~B., Bergin, E.~A., \& Plume, R.\ 2004, \apj, 613, 355
\bibitem[Kirk et al.(2013)]{Kirketal2013}
     Kirk, H., Myers, P.~C., Bourke, T.~L., et al.\ 2013, \apj, 766, 115
\bibitem[Klessen \& Hennebelle(2010)]{KlessenHennebelle2010}
     Klessen, R.~S., \& Hennebelle, P.\ 2010, \aap, 520, A17
\bibitem[Koyama \& Inutsuka(2000)]{KoyamaInutsuka2000}
     Koyama, H., \& Inutsuka, S.-I.\ 2000, \apj, 532, 980
\bibitem[Kudoh et al.(2007)]{Kudohetal2007}
     Kudoh, T., Basu, S., Ogata, Y., \& Yabe, T.\ 2007, \mnras, 380, 499
\bibitem[Kudoh \& Basu(2008)]{KudohBasu2008}
     Kudoh, T., \& Basu, S.\ 2008, \apjl, 679, L97
\bibitem[Kudoh \& Basu(2011)]{KudohBasu2011}
     Kudoh, T., \& Basu, S.\ 2011, \apj, 728, 123
\bibitem[Ledoux(1951)]{Ledoux1951}
     Ledoux, P.\ 1951, Annales d'Astrophysique, 14, 438
\bibitem[Li et al.(2004)]{Li_etal2004} Li, P.~S., Norman, M.~L., Mac Low, M.-M., 
     \& Heitsch, F.\ 2004, \apj, 605, 800
\bibitem[Li et al.(2013)]{Li_etal_2013} Li, H.-b., Fang, M., Henning, T.,
     \& Kainulainen, J.\ 2013, \mnras, 2585
\bibitem[Liu et al.(2012)]{Liuetal2012} Liu, H.~B., Jim{\'e}nez-Serra, I.,
     Ho, P.~T.~P., et al.\ 2012, \apj, 756, 10
\bibitem[Lindeberg(1998)]{Lindeberg1998}
     Lindeberg, T.\ 1998, International Journal of Computer Vision, 30, 117
\bibitem[Men'shchikov et al.(2010)]{Menshchikovetal2010}
     Men'shchikov, A., Andr\'e, P., Didelon, P., et al. 2010, A\&A, 518, L103+
\bibitem[Molinari et al.(2011)]{Molinarietal2011} Molinari, S., Bally, J., Noriega-Crespo, A.,
     et al.\ 2011, \apjl, 735, L33
\bibitem[Myers(1983)]{Myers1983}
     Myers, P.~C.\ 1983, \apj, 270, 105
\bibitem[Myers(2009)]{Myers2009}
     Myers, P. C. 2009, \apj, 700, 1609
\bibitem[Miyama et al.(1987)]{Miyamaetal1987}
     Miyama, S.~M., Narita, S., \& Hayashi, C.\ 1987, Progress of
     Theoretical Physics, 78, 1273
\bibitem[Mizuno et al.(1995)]{Mizunoetal1995}
     Mizuno, A., Onishi, T., Yonekura, Y., Nagahama, T., Ogawa, H.,
     \& Fukui, Y. 1995, ApJ, 445, 161
\bibitem[Nagai et al.(1998)]{Nagaietal1998}
     Nagai, T., Inutsuka, S.-I., \& Miyama, S.~M.\ 1998, \apj, 506, 306
\bibitem[Nakajima \& Hanawa(1996)]{NakajimaHanawa1996}
     Nakajima, Y., \& Hanawa, T.\ 1996, \apj, 467, 321
\bibitem[Nakamura \& Li(2008)]{Nakamura&Li2008} Nakamura, F., \& Li, Z.-Y.\ 
     2008, \apj, 687, 354 
\bibitem[Nakano \& Nakamura(1978)]{NakanoNakamura1978}
     Nakano, T., \& Nakamura, T.\ 1978, \pasj, 30, 671
\bibitem[Ostriker(1964)]{Ostriker1964}
     Ostriker, J.\ 1964, \apj, 140, 1056
\bibitem[Padoan \& Nordlund(2002)]{PadoanNordlund2002}
     Padoan, P., \& Nordlund, {\AA}.\ 2002, \apj, 576, 870
\bibitem[Palmeirim et al.(2013)]{Palmeirimetal2013}
     Palmeirim, P., Andr{\'e}, P., Kirk, J., et al.\ 2013, \aap, 550, A38
\bibitem[Peretto et al.(2012)]{Perettoetal2013}
     Peretto, N., Andr{\'e}, P., K{\"o}nyves, V., et al.\ 2012, \aap, 541, A63
\bibitem[Pon et al.(2012)]{Ponetal2012}
     Pon, A., Toal{\'a}, J.~A., Johnstone, D., et al.\ 2012, \apj, 756, 145
\bibitem[Schmid-Burgk(1967)]{Schmid-Burgk1967}
     Schmid-Burgk, J.\ 1967, \apj, 149, 727
\bibitem[Schneider \& Elmegreen(1979)]{SchneiderElmegreen1979}
     Schneider, S., \& Elmegreen, B. 1979, ApJS, 41, 87
\bibitem[Schneider et al.(2012)]{Schneideretal2012}
     Schneider, N., Csengeri, T., Hennemann, M., et al.\ 2012, \aap, 540, L11
\bibitem[Smith et al.(2012)]{Smithetal2012}
     Smith, R.~J., Shetty, R., Stutz, A.~M., \& Klessen, R.~S.\ 2012, \apj, 750, 64
\bibitem[Smith et al.(2013)]{Smithetal2013}
     Smith, R.~J., Shetty, R., Beuther, H., Klessen, R.~S., \& Bonnell, I.~A.\ 2013, \apj, 771, 24
\bibitem[Spitzer(1942)]{Spitzer1942}
     Spitzer, L., Jr. 1942, \apj, 95, 329
\bibitem[Tafalla et al.(2004)]{Tafallaetal2004}
     Tafalla, M., Myers, P.~C., Caselli, P., \& Walmsley, C.~M.\ 2004, \aap, 416, 191
\bibitem[Toal{\'a} et al.(2012)]{Toalaetal2012}
     Toal{\'a}, J.~A., V{\'a}zquez-Semadeni, E., \& G{\'o}mez, G.~C.\ 2012, \apj, 744, 190
\bibitem[Truelove et al.(1997)]{Trueloveetal1997}
     Truelove, J.~K., Klein, R.~I., McKee, C.~F., et al.\ 1997, \apjl, 489, L179
\bibitem[Umekawa et al.(1999)]{Umekawaetal1999}
     Umekawa, M., Matsumoto, R., Miyaji, S., \& Yoshida, T.\ 1999, \pasj, 51, 625
\bibitem[Van Loo et al.(2006)]{VanLooetal2006}
     van Loo S., Falle S.~A.~E.~G., Hartquist T.~W., 2006, \mnras, 370, 975
\bibitem[van Loo et al.(2007)]{VanLooetal2007}
     van Loo, S., Falle, S.~A.~E.~G., Hartquist, T.~W., \& Moore, T.~J.~T.\ 2007,
     \aap, 471, 213
\bibitem[van Loo et al.(2010)]{VanLooetal2010}
     van Loo, S., Falle, S.~A.~E.~G., \& Hartquist, T.~W.\ 2010, \mnras, 406, 1260
\bibitem[V{\'a}zquez-Semadeni et al.(2003)]{Vazquezetal2003}
     V{\'a}zquez-Semadeni, E., Ballesteros-Paredes, J., \& Klessen,
     R.~S.\ 2003, \apjl, 585, L131
\bibitem[Ysard et al.(2013)]{Ysardetal2013}
     Ysard, N., Abergel, A., Ristorcelli, I., et al.\ 2013, arXiv:1309.6489
\end{thebibliography}
\end{document}